\documentclass[lettersize,journal,twoside]{IEEEtran}
\usepackage{amsmath,amsfonts}
\usepackage{algorithmic}
\usepackage{algorithm}
\usepackage{array}
\usepackage[caption=false,font=normalsize,labelfont=sf,textfont=sf]{subfig}
\usepackage{textcomp}
\usepackage{stfloats}
\usepackage{url}
\usepackage{verbatim}
\usepackage{graphicx}
\usepackage{cite}
\usepackage{cite}
\usepackage{amsmath,amssymb,amsfonts}
\usepackage{algorithmic}
\usepackage{graphicx}
\usepackage{algorithm,algorithmic}
\usepackage[colorlinks=true,linkcolor=blue, allcolors=blue]{hyperref}
\hypersetup{colorlinks,
}
\usepackage{enumitem} 
\usepackage{textcomp}
\usepackage{multirow}
\usepackage{amsthm}
\usepackage{xcolor}
\usepackage{svg}
\usepackage{float}
\usepackage{pifont}
\usepackage{float}
\usepackage{graphicx}
\usepackage{tabularx}
\usepackage{tikz}
\usetikzlibrary{svg.path}

\definecolor{orcidlogocol}{HTML}{A6CE39}
\tikzset{
  orcidlogo/.pic={
    \fill[orcidlogocol] svg{M256,128c0,70.7-57.3,128-128,128C57.3,256,0,198.7,0,128C0,57.3,57.3,0,128,0C198.7,0,256,57.3,256,128z};
    \fill[white] svg{M86.3,186.2H70.9V79.1h15.4v48.4V186.2z}
                 svg{M108.9,79.1h41.6c39.6,0,57,28.3,57,53.6c0,27.5-21.5,53.6-56.8,53.6h-41.8V79.1z M124.3,172.4h24.5c34.9,0,42.9-26.5,42.9-39.7c0-21.5-13.7-39.7-43.7-39.7h-23.7V172.4z}
                 svg{M88.7,56.8c0,5.5-4.5,10.1-10.1,10.1c-5.6,0-10.1-4.6-10.1-10.1c0-5.6,4.5-10.1,10.1-10.1C84.2,46.7,88.7,51.3,88.7,56.8z};
  }
}

\newcommand\orcidicon[1]{\href{https://orcid.org/#1}{\mbox{\scalerel*{
\begin{tikzpicture}[yscale=-1,transform shape]
\pic{orcidlogo};
\end{tikzpicture}
}{|}}}}

\def\BibTeX{{\rm B\kern-.05em{\sc i\kern-.025em b}\kern-.08em
    T\kern-.1667em\lower.7ex\hbox{E}\kern-.125emX}}

\begin{document}
\title{A Data-Driven Framework for Online Mitigation of False Data Injection Signals in Networked Control Systems}
\author{Mohammadamin Lari
\thanks{I acknowledge the support of the Natural Sciences and Engineering Research Council of Canada (NSERC).}
\thanks{The author is with the School of Engineering, The University of British Columbia, Kelowna, BC V1V 1V7, Canada (e-mail: mohammadamin.lari@ubc.ca).}}


\maketitle

\begin{abstract}
This paper introduces a novel two-stage framework for online mitigation of False Data Injection (FDI) signals to improve the resiliency of Networked Control Systems (NCSs) and ensure their safe operation in the presence of malicious activities. The first stage involves meta learning to select a base time series forecasting model within a stacked ensemble learning architecture. This is achieved by converting time series data into scalograms using continuous wavelet transform, which are then split into image frames to generate a scalo-temporal representation of the data and to distinguish between different complexity levels of time series data based on an entropy metric using a convolutional neural network. In the second stage, the selected model mitigates false data injection signals in real-time. The proposed framework's effectiveness is demonstrated through rigorous simulations involving the formation control of differential drive mobile robots. By addressing the security challenges in NCSs, this framework offers a promising approach to maintaining system integrity and ensuring operational safety. 
\end{abstract}


\section{Introduction} \label{sec1}
\IEEEPARstart{N}{etworked} Control Systems (NCSs) are created when a network is used to close the control loop, which consists of components such as sensors to measure process variables, controllers to generate commands for achieving a desired goal, and actuators to execute these commands. NCSs are widely used in industrial sites, smart grids and intelligent transportation systems, to manage these infrastructures. In these systems, sensors, controllers, and actuators are interconnected via a network, which facilitates the exchange of data between components that are physically distributed and continuous monitoring of the operational data. However, utilizing networks makes system vulnerable to malicious cyberattacks. In the last decade, multiple incidents of cyberattacks with drastic impacts against NCSs have been reported \cite{hemsley2018history}. 

Depending on the amount of an adversary's access to three key resources, i.e., system model knowledge, disclosure resources, and disruption resources, various cyberattack scenarios are studied in the literature \cite{TEIXEIRA2015135}. 
The transmission of data through communication links is blocked in Denial of Service (DoS) attacks by overwhelming the targeted network with a flood of traffic or requests \cite{10175196}. A combination of resources is needed to alter the transmitted data in some attack scenarios, such as deception attacks \cite{10041024}, which include forms like Replay and False Data Injection (FDI) \cite{9771390} attacks. The adversary can create more sophisticated attacks, such as covert attacks \cite{10262318} with access to all three key resources. Among the discussed cyberattack scenarios, FDIs are the most common. This is because executing FDI attacks generally requires minimal access to the mentioned resources, whereas designing replay and covert attacks requires greater access, which might be significantly more demanding in some cases. Consequently, this paper focuses on the FDI attack scenario.
In FDI attacks, the adversary alters sensor measurements or control signals with the aim of degrading the system performance or endangering its stability. 

To prevent any damage to NCSs and drastic financial losses, it is necessary to equip these systems with appropriate mechanisms, including predictive monitoring modules \cite{9740504} and resilient control algorithms \cite{8933052}. These mechanisms include detection, isolation, and mitigation phases using model-based approaches, data-driven approaches, or a combination of both \cite{ZHANG20211}. 
Model-based approaches for FDI detection depend on an understanding of the system to develop a parameterized model of the process \cite{9054955, 9857780}. 
To prevent malicious data from spreading throughout the system, it is necessary to isolate the compromised subsystems from the network. Several algorithms are proposed for this purpose, including monitoring the reputation of agents \cite{6736104}, and weighted mean sub-sequence reduced algorithm \cite{6481629}. An additional crucial aspect in control system design is the mitigating FDI attacks to ensure the safe operation of NCSs, which involves reconstructing the FDI attack and compensating for its impacts in real-time \cite{9474579, 9745491, 9540250, 9697096}. 

Alternatively, significant research efforts are being directed towards developing data-driven approaches for real-time detection, isolation, and mitigation of FDI attacks. Developing data-driven methods is beneficial due to utilizing the system's historical operational data instead of having prior knowledge of the system's dynamics. Detecting FDI attacks using data-driven methods is studied in \cite{9319658, 7926429, en12112209}. The literature contains a limited number of studies focused on mitigating FDI attacks through data-driven approaches. An FDI mitigation framework using subspace identification method is presented in \cite{10071718}. However, the proposed method is restricted to linear systems, which may not be realistic in many cases. Moreover, it lacks incorporating both time and frequency domain information in the framework, which may result in missing some of the features necessary for effectively mitigating FDI attacks. Long Short-Term Memory (LSTM) networks are used for mitigating FDI attacks in \cite{9352502}. Similarly, this framework is developed for systems with linear models, and do not incorporate both time and frequency domain information for mitigating FDI attacks. Additionally, this framework is restricted by the assumption that attack signals are whether constant or periodic. A combination of Leuenberger observer and artificial neural network is proposed in \cite{8855088} for mitigation of FDIs, but this framework requires prior knowledge of the system model for designing the observer. Moreover, this framework is limited to systems with linear models, does not incorporate both time and frequency domain features, and assumes the FDI attack is periodic. An adaptive neuro-fuzzy inference system for FDI mitigation is proposed in \cite{en15228539}, but the proposed method has a high computational cost and scalability issues due to exponential growth of the number of fuzzy rules, and lacks using both time and frequency domain knowledge for FDI mitigation. An attention-based temporal convolutional denoising auto-encoder for mitigating FDI attacks is developed in \cite{RAGHUVAMSI2023112565}. However, the proposed framework may not be accurate or robust in some cases due to using a single model for all attack scenarios, and lacks incorporating both time and frequency domain information for FDI mitigation.
\begin{table}[t]
\caption{Comparison of proposed framework with existing literature}
	\begin{center}
	\label{tab3}
\begin{tabular}{|c|c|c|c|c|c|}
\hline
Shortcoming                & \ \ (i) \ \  & \ \ (ii) \ \ & \ \ (iii) \ \ &\ \ (iv) \ \ & \ \ (v) \ \ \\ \hline
 Framework \cite{10071718}  & \ding{55}           & \ding{51}           & \ding{55}           & \ding{55}           & \ding{51} \\ 
 Framework \cite{9352502}      & \ding{55}   & \ding{51}           & \ding{55}           & \ding{55}           & \ding{55} \\
 Framework \cite{8855088}      & \ding{55}   & \ding{55}           & \ding{55}           & \ding{55}           & \ding{55} \\
 Framework \cite{en15228539}      & \ding{55}   & \ding{51}           & \ding{51}           & \ding{55}           & \ding{51} \\
 Framework \cite{RAGHUVAMSI2023112565}      & \ding{55}   & \ding{51}           & \ding{51}           & \ding{55}           & \ding{51} \\
Proposed Framework & \ding{51}   & \ding{51}           & \ding{51}           & \ding{51}           & \ding{51} \\ \hline
\end{tabular}
\end{center}
\end{table}
It is worth noting that all of the existing data-driven approaches for mitigating FDI attacks rely on using a single model, either to approximate the behavior of an individual sensor within the system or to approximate the dynamics of the entire system for all of FDI attack scenarios. This leads to performance degradation if the model is too simple to accurately handle complex data, or results in excessive computational burden if the model is overly complex. An alternative approach is using an ensemble of models \cite{10018517}. A popular ensemble learning technique is stacking, where multiple base models that are skillful in a particular task are trained individually, and a model selection module, usually referred to as meta learner, is responsible for dynamically selecting the best base model. Stacked architectures are typically highly accurate, which is a key reason for their widespread use. Additionally, stacked architectures benefit from improved diversity by employing different algorithms to train the base models \cite{9893798}. A stacked architecture is introduced in \cite{WANG2022100542} for detecting FDI attacks by combining the outputs of various models. However, the proposed framework lacks dynamic model selection and simply averages the predictions of the base models, which is not an efficient strategy. 

The existing literature for mitigating FDI attacks has several shortcomings: (i) absence of ensemble learning-based frameworks, leading to performance degradation, either due to the model being too simple to accurately handle complex data, or causing excessive computational burden if the model is overly complex, (ii) requirement of prior system model knowledge to obtain a parameterized model of the process, making the developed frameworks impractical in some cases, where obtaining an accurate model may be difficult or even impossible, (iii) only being applicable to systems with a linear model, which may not be practical because many any real-world systems exhibit nonlinear behaviors, (iv) lack of incorporating both time and frequency domain information into mitigation algorithms, which may result in failing to leverage crucial features necessary for effectively mitigating FDI attacks, and (v) prevalence of conservative assumptions, particularly regarding periodic attack signals, limiting the generalization property of the solutions to various attack scenarios. This paper addresses the literature shortcomings, through a computationally efficient two-stage data-driven framework for online mitigation of FDI attacks, which is compared against the existing literature in Table \ref{tab3}, with the following contributions: 
\begin{enumerate}
    \item A novel two-stage data-driven framework for online mitigation of FDI attacks in NCSs is proposed, where the first stage performs model selection for dynamically selecting a base time series forecasting model for FDI mitigation purposes through classification of complexity level of time series data, and the second stage mitigates the FDI attack using the selected model in real-time.
    \item Distinct characteristics of time series data are identified by performing time-frequency analysis and incorporating frequency domain features into data, resulting in scalo-temporal feature extraction. This is achieved by applying continuous wavelet transform on time series data to generate scalograms and transforming them into multi-channel scalogram image frames. The channels within an image frame represent shifts in the time domain to capture the dynamic spectrum of the time series data.
    \item The complexity of the time series data segments, corresponding to the multi-channel scalogram image frames, is analyzed using an entropy metric by the proposed Optimally Improved Permutation Entropy (OIPE) algorithm. This involves symbolizing the ordinal patterns of the time series data and selecting the optimal number of quantization levels via a cross-validation procedure. 
    \item Dynamic selection of the base time series forecasting model for mitigating the FDI attack is performed by leveraging the generated multi-channel scalogram image frames, and their corresponding complexity levels. This framework is the first attempt to use a time-frequency domain approach for dynamic model selection in stacked ensemble learning architectures to prevent the computational inefficiency caused by utilizing an individual overly complex model.
\end{enumerate}

The rest of this paper is organized as follows: Section \ref{sec2} presents the problem formulation and required background knowledge as well as Stage I of the proposed framework. In section \ref{sec3}, a detailed description of Stage II of the proposed framework for online mitigation of FDI attacks is included. Section \ref{sec4} is allocated to introducing the simulation setup, discussing the results, and conducting comparative studies. In the end, section \ref{sec5} concludes the paper and proposes possible future research directions to extend this work. 

\section{Stage I - Meta Learning Based on Classification of Data in Time-Frequency Domain} \label{sec2}
\subsection{Problem Formulation}
\begin{figure}[t] 
	\centerline{\includegraphics[width=7cm]{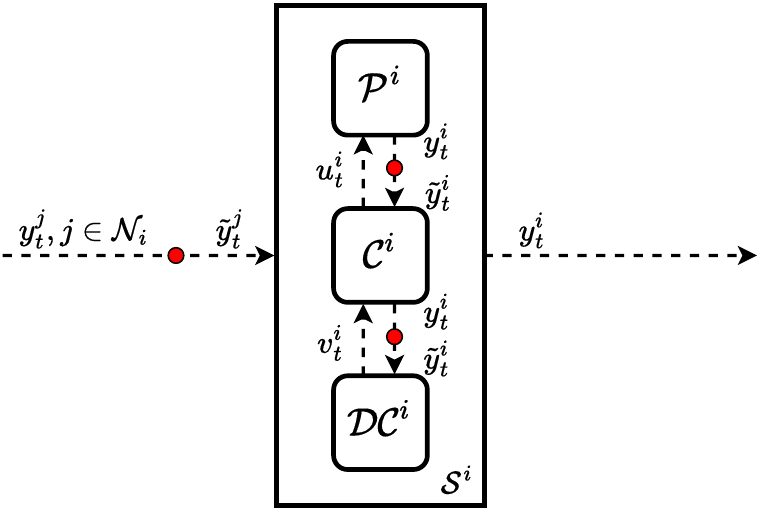}}
	\caption{Diagram representation of the subsystems in a NCS with hierarchical structure and the security vulnerabilities (red circles)}
	\label{fig12}
\end{figure}

Consider a network of $N$ subsystems with hierarchical structure \cite{TIPSUWAN20031099}, as depicted in Fig. \ref{fig12}. The $i^{\text{th}}$ subsystem, denoted by $\mathcal{S}^i$, is comprised of a plant $\mathcal{P}^i$, a high level distributed controller $\mathcal{DC}^i$ to generate the reference signal $v^i_t$ for the subsystem, and a low level local controller $\mathcal{C}^i$ to regulate the behavior of the plant by receiving sensor measurement $y^i_t$ and reference signal $v^i_t$ from the plant and the distributed controller, respectively, and generating the control command $u^i_t$. The reference signal $v^i_t$ is generated by the distributed controller using local sensor measurements $y^i_t$ and received sensor measurements from neighboring subsystems $\mathcal{N}_i \subseteq \mathcal{N}=\{1,2, ... , N\}$ through a communication network. As shown in Fig. \ref{fig12}, the sensor measurements are vulnerable to FDI attacks. For the rest of this article, it is assumed that the sequence of measurements of a single sensor in the NCS are represented as time series data and denoted by the general notation $Y$ as follows:
\begin{equation} \label{eq2.1}
    Y = \left \{ y_t, \quad t\in I \right \}
\end{equation}
where $y_t$ represents a measurement of the sensor at time instant $t$, and $I$ is the interval of the complete time series data. It is assumed that the attacker has access to the required resources and can execute an FDI attack, denoted by $a_t$ on the sensor measurement $y_t$ to generate a compromised sensor measurement $\tilde y_t$ using the following equation:
\begin{equation} \label{eq2}
	\tilde y_t = y_t + a_t
\end{equation}
Consequently, the time series data may include the following sequence of compromised sensor measurements:
\begin{equation} \label{eq202}
    \tilde Y = \left \{\tilde y_t, \quad t\in \tilde I \right \}
\end{equation}
where $\tilde I\subseteq I$ is the interval that $y_t$ is compromised. 

The objective of this article is developing a data-driven framework to obtain the value of the injected signal on sensor measurement $y_t$, and mitigate its effects in real-time to ensure the safe operation of subsystems in the network, even in the presence of FDI attacks. The FDI attack is reconstructed using the error between the predicted and actual sensor measurements. Then, the obtained FDI attack reconstruction is used to adjust the control commands in real-time and mitigate the FDI attack impacts from the target subsystem.

\subsection{Stacked Ensemble Learning Architecture}
Stacked ensemble learning architectures integrate predictions from various base models. Using a stacked architecture offers the advantage of providing more accurate predictions compared to any of the individual base models. In stacking, multiple base models are trained individually on the dataset. However, using multiple base models simultaneously is not computationally efficient. Consequently, a model selection module is required to choose a subset of the base models and reduce the computational burden of the framework. In this paper, model selection is performed by a meta learner, which extracts time-frequency domain features from time series data and assesses complexity level using an entropy-based metric.

\subsection{Phase I: Scalo-Temporal Feature Extraction using Continuous Wavelet Transform}
Time-frequency domain analysis offers significant advantages over pure time or frequency domain methods by providing a more complete representation of signals. Unlike time domain analysis, which focuses on signal evolution over time, or frequency domain analysis, which shows the spectral content without temporal detail, time-frequency analysis captures both. This allows for precise detection of changes in frequency content over time, making it ideal for analyzing complex time series data, such as FDI attacks. It improves the ability to identify transient behaviors and subtle variations that might be missed using raw time series data. There are various algorithms for time-frequency domain analysis, such as short-time Fourier transform (STFT) and continuous wavelet transform. STFT splits the signal into multiple segments of equal length using a sliding window and applies Fourier transform on each segment. Consequently, STFT allows for determining where in the signal the frequency content has occurred. The main issue with this algorithm is that it encounters the theoretical constraints of the Fourier transform, known as the uncertainty principle \cite{folland1997uncertainty}. Reducing the window size enhances our understanding of when a frequency appears in the signal but reduces our knowledge of the precise frequency value. Conversely, increasing the window size improves our knowledge of the frequency value but provides less information about its occurrence in time. Since the window length is constant in the STFT algorithm, information will be lost in both time and frequency domains.  

A more effective method for analysing signals with a dynamic frequency spectrum is applying a wavelet transform on the signal. Wavelet transform decomposes a signal into scaled and shifted versions of the original wavelet, known as the mother wavelet \cite{graps1995introduction}. This ability allows for the examination of the function across various time and frequency scales, making wavelets highly effective tools for processing time series data. There are various families of wavelets, where the choice of wavelet family is dictated by the signal characteristics and the nature of the application \cite{gonccalves2022improved}. Wavelet transform uses a short window length at high frequencies and a long window length at low frequencies \cite{91217}. Consequently, Wavelet transform offers high resolution in both the frequency and time domains and utilizes wavelets instead of window functions. There are two types of Wavelet transform: Discrete Wavelet Transform (DWT) and Continuous Wavelet Transform (CWT). In CWT, a signal is represented by continuously adjusting the translation and scale parameters of the wavelets. This process provides a time-frequency representation of the signal, offering precise localization in both time and frequency domains \cite{9315960}.

To fully extract the frequency domain features of the time series data $Y$ over time, it is split into $N_m$ segments, referred to as frames, using a window length $M$ and overlap $L$, using the following equation:
\begin{equation} \label{eq4}
    N_m=\left \lfloor \frac{N_y-L}{M-L} \right \rfloor + 1
\end{equation}
where $\left \lfloor \cdot \right \rfloor$ denotes the floor function, $N_y$ is the number of data points in time series $Y$, as defined in (\ref{eq2.1}), and $M$ is determined as follows: 
\begin{equation} \label{eq5}
    M=\lambda f_s
\end{equation}
where $\lambda$ is a scaling factor that depends on the hardware and has to be chosen appropriately to allow for real-time processing of the time series data, and $f_s$ is the sampling frequency. Then, a 2D CWT matrix representation of the $i^{\text{th}}$ frame $W^{(i)} \in \mathbb{R}^{S\times M}$ is obtained as follows:
\begin{equation} \label{eq6}
    W^{(i)}=\begin{bmatrix}
w^{(i)}(1,1) & \cdots  & w^{(i)}(1,M)\\ 
\vdots  & \ddots  & \vdots \\ 
 w^{(i)}(S,1)& \cdots  & w^{(i)}(S,M) 
\end{bmatrix}
\end{equation}
where $ i\in \left \{ 1,2,\cdots ,N_m \right \}$, $S$ is the number of scales, and each element of this matrix, denoted by $w^{(i)}(s ,m)$, is calculated using CWT, defined by the following equation:
\begin{equation} \label{eq7}
     w^{(i)}(s ,m)=\frac{1}{\sqrt{s}}\sum_{t=0}^{s-1}y_{m+t}\psi_{ts^{-1}}e^{-j2\pi t}
\end{equation}
where $s$ is the wavelet scale and can take on $S$ different values within the range $\left [ S_{min}, S_{max} \right ]$, $m \in \left \{ 1,\cdots M \right \} $ is the shifting value, and $\psi_t$ is the window function that is selected from one of the wavelet families \cite{mateo2020bridging}. $S_{min}$ and $S_{max}$ are the minimum and maximum scales of the CWT, which are determined based on the energy spread of the wavelet in frequency and time. Among the existing wavelet families, Morse wavelet is chosen in this paper due to providing a satisfactory resolution in the frequency domain. The Morse wavelet is defined as follows:
\begin{equation} \label{eq8}
    \psi_t\Leftrightarrow \psi _{\beta ,\gamma }(\omega)=U(\omega)a_{\beta ,\gamma}\omega ^{\beta }e^{-\omega ^{\gamma }}
\end{equation}
where $\Leftrightarrow$ denotes a Fourier transform pair, $U(\cdot)$ is the unit step function, $a_{\beta ,\gamma}$ is a normalization constant, $\beta$ controls the wavelet decay in time domain, and $\gamma$ controls the frequency domain decay of the wavelet \cite{5508620}. Then, $W^{(i)}$ is visualized through a scalogram frame, denoted by $SC^{(i)} \in \mathbb{R}^{S\times M}$ through mapping the amplitude of CWT matrix elements onto the time-scale plane, using different colors to represent variations in the intensity of the wavelet transform amplitudes across different scales or frequencies and time points. Scalogram represents the local time-frequency energy density and is obtained using the following equation \cite{7292231}: 
\begin{equation} \label{eq9}
    SC^{(i)}=\begin{bmatrix}
\left |w^{(i)}(1,1)  \right |^{2} & \cdots  & \left |w^{(i)}(1,M)  \right |^{2}\\ 
\vdots  & \ddots  & \vdots \\ 
 \left |w^{(i)}(S,1)  \right |^{2}& \cdots  & \left |w^{(i)}(S,M)  \right |^{2} 
\end{bmatrix}
\end{equation}
where $ i\in \left \{ 1,2,\cdots ,N_m \right \}$, and $\left| \cdot \right|$ denotes the absolute value function.

Extracting scalo-temporal features from raw time series data using CWT and visualization of the CWT matrix using scalogram frames provides a high resolution description of data in both time and frequency domains. To further process the time series data, the scalogram frames are represented as image frames, denoted by $P^{(i)}_{o}$ for the $i^{\text{th}}$ scalogram frame with dimensions ${w} \times {h} \times c$, where ${w}$ and ${h}$ denote the image's width and height, respectively, and $c$ denotes the number of channels, which is equal to three because scalogram frames are originally visualized as RGB images. This representation enables time-frequency domain analysis through the unique patterns present in each image. However, there is a trade-off between the frame length parameter $M$ and the effectiveness of the time-frequency domain analysis. While increasing the frame length enhances the quality of the analysis, it also raises the computational complexity of the algorithm, which is undesirable for real-time data processing. To address this challenge, the image frames need to capture variations over time, while maintaining the frame length as short as possible. The solution to this challenge involves two steps: 
\begin{enumerate}
    \item The images from previous step $P^{(i)}_{o}$ are converted to grayscale images to reduce the number of channels to one. Converting image frames to grayscale allows for stacking sequential image frames channel-wise, and using different channels within an image as an indication of time. The conversion to grayscale is performed using Luminosity method, as follows \cite{10132453}:
        \begin{equation} \label{eq12}
        \begin{aligned}
        P^{(i)}_{gr}(p_x,p_y) &= 0.299P^{(i,1)}_{o}(p_{x},p_{y}) \\
                       & + 0.587P^{(i,2)}_{o}(p_{x},p_{y}) \\
                       & + 0.114P^{(i,3)}_{o}(p_{x},p_{y})
        \end{aligned}
        \end{equation}
    \item ${c}'$ sequential image frames are stacked together to generate a ${w} \times {h} \times {c}'$ multi-channel image. The parameter ${c}'$ is determined based on the sampling frequency $f_s$ as follows:
    \begin{equation} \label{eq13}
        {c}'=\kappa f_s
    \end{equation}
    where $\kappa$ is a scaling factor that depends on the bandwidth of the time series data.
\end{enumerate} 

The multi-channel images $P^{(i)}_{gr}$ are then resized to have dimensions $\zeta \times \zeta \times {c}'$, where $\zeta$ depends on the frame length $M$. To maintain low computational complexity, the image size $\zeta$ should be chosen as the smallest possible value that preserves distinguishable scalogram features. Reducing the computational load of the algorithm without sacrificing the quality of the input features for accurate analysis is crucial for real-time processing of data. The resized multi-channel images are denoted by $P^{(i)}_{g}$. A pixel in $P^{(i)}_{g}$ with coordinates $(p_x,p_y)$ is obtained from the following
pixel in $P^{(i)}_{gr}$:
\begin{equation} \label{eq14}
    P^{(i)}_{g}(p_x,p_y) = P^{(i)}_{gr}(\left \lfloor \frac{p_x.w}{\zeta}\right \rfloor,\left \lfloor \frac{p_y.h}{\zeta}\right \rfloor) 
\end{equation}
Consequently, the $j^{\text{th}}$ channel of the resized multi-channel image is defined as follows:
    \begin{equation} \label{eq15}
        P^{(i,j)}_{g}=\begin{bmatrix}
    (p_0,p_0) & \cdots  &(p_0 ,p_\zeta ) \\ 
    \vdots  & \ddots  &\vdots  \\ 
    (p_\zeta ,p_0) & \cdots  &(p_\zeta,p_\zeta) 
    \end{bmatrix}
    \end{equation}
    where $j\in \left \{ 1, 2, \cdots ,{c}' \right \}$. Given the description of generating resized multi-channel scalogram image frames from raw time series data, the scalo-temporal feature extraction phase of the proposed framework is summarized in Algorithm \ref{alg4}.
 
\begin{algorithm}[t]
\caption{Phase I: Scalo-Temporal Feature Extraction using CWT}\label{alg4}
\textbf{Input:} $ Y $ \\ \vspace{0.2cm}\hspace{-0.11 cm}
\textbf{Output:} $P^{(i)}_{g}$ \\
\vspace{-0.5cm}
\begin{algorithmic}[1]
    \STATE Define time series length $N_y$, overlap $L$, sampling frequency $f_s$, image size $\zeta$, and scaling factors $\lambda$ and $\kappa$
    \STATE Calculate scalogram frame length $M$ using (\ref{eq5})
    \STATE Calculate number of frames $N_m$ using (\ref{eq4})
    \FOR{$ ( 1\le i \le N_m) $}
        \STATE Calculate CWT matrix $W^{(i)}$ using (\ref{eq6})-(\ref{eq8})
        \STATE Generate scalogram frame $SC^{(i)}$ using (\ref{eq9})
        \STATE Generate grayscale scalogram frame images $P^{(i)}_{gr}$ using (\ref{eq12})
        \STATE Compute number of channels ${c}'$ using (\ref{eq13})
        \STATE Generate the resized multi-channel images $P^{(i)}_{g}$ using (\ref{eq14}) and (\ref{eq15})
    \ENDFOR
\end{algorithmic}
\end{algorithm}

\subsection{Phase II: Complexity Analysis using Entropy Metric}
Entropy is a highly effective metric for assessing the complexity of time series data. It quantifies complexity in terms of the degree of disorder. A higher entropy value indicates a greater level of complexity, suggesting that the signal is more disordered \cite{AZAMI201628}. The calculation of entropy involves examining the temporal sequence of the data points and identifying the relative frequency of distinct patterns that occur. These patterns, known as ordinal patterns, represent the arrangement of values within a specified segment of the time series. By analyzing these patterns, entropy provides a comprehensive picture of time series' complexity. There are various entropy algorithms in the literature, including Approximate Entropy, Sample Entropy, Fuzzy Entropy, and Permutation Entropy \cite{10179861}. Among these, the Permutation Entropy algorithm is well-suited for analyzing real-world data due to its simplicity and fast computation. However, the Permutation Entropy algorithm has several limitations: 1) It does not account for the amplitude information of the signals, 2) It is sensitive to noise in real-world data, and 3) The entropy value is influenced when equal values occur in the time series. The Improved Permutation Entropy (IPE) algorithm addresses these limitations using symbolization of ordinal patterns, but the outcome is highly dependent on selection of the number of quantization levels \cite{chen2019improved}. To overcome this issue, we propose to adjust this parameter through a $K$-fold cross-validation procedure, which leads to the OIPE algorithm.

To calculate the corresponding entropy for $P^{(i)}_{g}$, it is binarized using the threshold $\tau _b = (\text{max}\{P^{(i)}_{g}\}-\text{min}\{P^{(i)}_{g}\})/2 $ as follows:
\begin{equation} \label{eq17}
    P_b^{(i,j)}(p_x,p_y)=\left\{
    \begin{array}{cl}
 0, &   P_g^{(i,j)}(p_x,p_y) \leq \tau _b\\ 
 1, & P_g^{(i,j)}(p_x,p_y) > \tau _b\\ 
\end{array}
\right.
\end{equation}
where $P_b^{(i,j)}(p_x,p_y)$ is the pixel with coordinate $(p_x,p_y)$ in the $j^{\text{th}}$ channel of $P^{(i)}_{b}$ with dimensions $\zeta \times \zeta \times {c}',\ j\in \left \{ 1, 2, \cdots ,{c}' \right \} $. 
Based on the method proposed in \cite{1211511}, a linear scanning method is performed to represent the $j^{\text{th}}$ channel of $P_b^{(i)}$ as a time series, denoted by $T_b^{(i,j)}\in \mathbb{R}^{ \zeta}$, using the following equation:
\begin{equation} \label{eq18}
    T_b^{(i,j)}(p_x) = \sum_{p_y=1}^{\zeta}P_b^{(i,j)}(p_x,p_y)
\end{equation}
Then, the fused time series corresponding to $P^{(i)}_{g}$, denoted by $Y_F^{(i)}$, is calculated as follows:
\begin{equation} \label{eq19}
    Y_F^{(i)} = T_b^{(i,1)} \odot T_b^{(i,2)} \odot \cdots  \odot  T_b^{(i,{c}')}
\end{equation}
where $ \odot $ denotes the element-wise product. The fused time series contains richer information in both the time and frequency domains compared to the raw time series, resulting in a more accurate complexity analysis. To determine the complexity of $Y_F^{(i)}$, the IPE with an embedding dimension $D$ and time delay $\tau$ is calculated. The fused time series $Y_F^{(i)}$ with length $\zeta$ is split into segments of length $D$ to form the embedding vectors as follows:
\begin{equation} \label{eq20}
    e(j)=\begin{bmatrix}
e_j & e_{j+\tau} & \cdots  & e_{j+(D-1)\tau}
\end{bmatrix}
\end{equation}
where $ j\in \left \{ 1, 2, \cdots , N_D \right \}, N_D=\zeta-(D-1)\tau$. Then, the embedding vectors are used as a basis to form an embedding space, denoted by $E\in \mathbb{R}^{ N_D \times D}$ as follows:
\begin{equation} \label{eq21}
    E = \left [ e(1)^T \quad e(2)^T\quad \cdots \quad e(N_D)^T  \right ]^T
\end{equation}
Let $y_{min}$ and $y_{max}$ be the minimum and maximum value of the time series data $Y_F^{(i)}$, respectively, and $H$ denote the number of quantization levels, where $H\in \left \{ H_{min}, H_{min}+1, \cdots ,H_{max} \right \}$. Then, the first column of the embedding space is symbolized using Uniform Quantization (UQ) function, defined by the following equation:
\begin{equation} \label{eq22}
    UQ(\mu)=\left\{
    \begin{array}{cl}
0, & y_{min}\leq \mu < y_{min}+\Delta\\ 
1, & y_{min}+\Delta \leq \mu <y_{min}+2\Delta\\ 
\vdots & \vdots\\ 
H-1, & y_{max}-\Delta \leq \mu \leq y_{max}
\end{array}
\right.
\end{equation}
where $\mu$ is the $j^{\text{th}}$ element of $E(:,1),\ j\in \left \{ 1, 2, \cdots ,N_D \right \}$, and $\Delta=(y_{max}-y_{min})/H$. The UQ function maps the first column of the embedding space to a symbolized sequence, denoted by $Q(:,1)$, ranging from 0 to $H-1$. Then, the $k^{\text{th}}$ column of embedding space $E(:,k)$ is calculated using the following equation:
\begin{equation} \label{eq23}
    Q(j,k)=Q(j,1)+\left \lfloor \frac{(E(j,k)-E(j,1))}{\Delta} \right \rfloor
\end{equation}
where $k\in \left \{ 1, 2, \cdots ,D \right \}$. The resulting matrix $Q$ is known as the symbolic space, and the unique rows in this symbolic space are referred to as symbolic patterns, denoted by $\pi_i,\ i\in \left \{ 1, 2, \cdots ,H^D \right \}$. Then, the probability distribution of the symbolic patterns, denoted by $ p_i=N_{\pi_i}/ N_D$, is computed, where $N_{\pi_i}$ represents the number of occurrences of the symbolic pattern $\pi_i$ in the time series $Y_F^{(i)}$. Then, the normalized entropy value is calculated as follows:
\begin{equation} \label{eq24}
    En(Y_F^{(i)})=\frac{-\sum_{i=1}^{H^D}p_i\text{ln}(p_i)}{\text{ln}(H^D)}
\end{equation}

The above equation highlights that the number of quantization levels $H$, significantly influences the calculation of entropy, as it is crucial in the symbolization process. A higher $H$ value preserves more time series information during symbolization, while a lower $H$ value enhances robustness against noise. Selecting an appropriate $H$ depends on the characteristics of the time series, such as its signal-to-noise ratio. However, this information is typically unknown beforehand. Therefore, a $K$-fold cross-validation \cite{zhang1993model} procedure is applied to adjust the parameter $H$ by dividing the dataset into \( K \) folds. For each fold, \( K-1 \) folds are considered as training, and the remaining fold is considered as validation, and the entropy value is calculated for both subsets. This process is repeated \( K \) times, each time with a different fold used for validation, yielding \( K \) pairs of training and validation entropy values. The cross-validation error for parameter \( H \), denoted by $Err(H)$, is defined as the average difference in entropy between the training and validation sets using the following equation:
\begin{equation} \label{eq25}
    Err(H) = \frac{1}{K} \sum_{i=1}^{K} \left| En(Y_{tr, i}) - En(Y_{vl, i}) \right|
\end{equation}
where \( Y_{tr, i} \) and \( Y_{vl, i} \) are the training and validation sets, respectively, when the \( i^{\text{th}}\) fold is used for validation. This process is repeated for different values of \( H \), and the parameter value that yields the smallest average difference is selected as the optimal number of quantization levels, denoted by $H^*$ as follows:
\begin{equation} \label{eq26}
    H^*=\min_H{Err(H)}
\end{equation}
This procedure ensures that the chosen number of quantization levels is optimal, thereby improving the robustness and accuracy of the entropy calculation. Then, the corresponding entropy for $P^{(i)}_{g}$ is computed as follows:
\begin{equation} \label{eq27}
    En^{(i)}=En(Y_F^{(i)})   
\end{equation}

To partition the dataset into $r$ levels of complexity, $r-1$ threshold values, denoted by $\tau_{e}^k,\ k\in\left \{ 1,2,\cdots ,r-1 \right \}$ are defined. These threshold values are chosen by computing the entropy values for the validation dataset using the proposed OIPE algorithm. The complexity level corresponding to $P^{(i)}_{g}$ is determined using the following equation:
\begin{equation} \label{eq28}
    L_c^{(i)}=\left\{
    \begin{array}{cl}
     1, & 0\leq  En^{(i)} < \tau_{e}^1\\ 
     2, & \tau_{e}^1 \leq En^{(i)} < \tau_{e}^2\\ 
    \vdots & \vdots\\ 
     r, & \tau_{e}^{r-1} \leq En^{(i)} \leq 1
    \end{array}
    \right.
\end{equation}
where $r$ is determined based on the number of time series forecasting models in the stacked ensemble learning architecture, $n$, as follows:
\begin{equation} \label{eq29}
    r=\left \lfloor \rho n \right \rfloor
\end{equation}
where $\rho$  is a scaling factor that varies with the dataset and should be chosen to optimize the utilization of the hardware's computational resources.
Given the detailed description of the steps for complexity analysis using entropy metric, Algorithm \ref{alg5} summarizes the steps for determining the complexity level corresponding to multi-channel scalogram image frames.

\begin{algorithm}[t]
\caption{Complexity Analysis using Entropy Metric}\label{alg5}
\textbf{Input:} $P^{(i)}_{g}$ \\ \vspace{0.2cm}\hspace{-0.11 cm}
\textbf{Output:} $L_c^{(i)}$ \\
\vspace{-0.5cm}
\begin{algorithmic}[1]
    \STATE Define the number of frames $N_m$, number of folds $K$, embedding dimension $D$, time delay $\tau$, number of complexity levels $r$, and complexity thresholds $\tau_e^i, i\in\left \{ 1,2,\cdots ,r-1 \right \}$
    \FOR{$ ( 1\le i \le N_m) $}
        \STATE Generate $Y_F^{(i)}$ using (\ref{eq17})-(\ref{eq19})  
        \STATE Compute the embedding space $E$ using (\ref{eq20}) and (\ref{eq21})
        \FOR{$ ( H_{min} \le H \le H_{max} )$}
            \STATE Compute the symbolic space $Q$ using (\ref{eq22}) and (\ref{eq23})
            \STATE Compute the entropy $En(Y_F^{(i)})$ using (\ref{eq24})
            \STATE Divide the time series data into $K$ folds
            \FOR{$ ( 1 \le k \le K )$}
                \STATE Determine $Y_{tr,k}$ and $Y_{vl,k}$
                \STATE Compute $Err(H)$ using (\ref{eq25})
            \ENDFOR
        \ENDFOR
        \STATE Determine $H^*$ using (\ref{eq26})
        \STATE Generate the complexity level $L_c^{(i)}$ using (\ref{eq27})-(\ref{eq29})
    \ENDFOR
\end{algorithmic}
\end{algorithm}

\subsection{Phase III: FDI Complexity Classification using CNN}
Based on the scalo-temporal features extracted from time series data $Y$ and their corresponding complexity levels, as described in phases I and II, model selection can be performed to intelligently select a base time series forecasting model from the ensemble based on the complexity of the time series data, enhancing the computational efficiency of the prediction algorithm. Following the steps in the previous phases, the raw time series data is in the form of $D^{(i)}=\left \{ P^{(i)}_{g},\ L_c^{(i)} \right \},\ i\in\left \{ 1,2,\cdots ,N_m \right \}$. Then, a Convolutional Neural Network (CNN) is trained offline to learn the relation between scalo-temporal features and the complexity level of time series data. The model is trained in a supervised manner using dataset $D=\left \{ D^{(i)} \right \}_{i=1}^{N_m}$, which is partitioned into training and validation subsets. The trained model predicts the probability of an input data belonging to each of the complexity levels, denoted by $P^i$ for the $i^{\text{th}}$ class. Then, the normalized confidence scores are defined to ensure that the probabilities sum to one as follows:
\begin{equation} \label{eq33}
    C_{n}^{i}=\frac{ P^{i}}{\sum_{i=1}^{r}P^{i}}
\end{equation}
The predicted class for an input data is computed as the class index with maximum normalized confidence score as follows:
\begin{equation} \label{eq34}
    \hat L_c=\underset{i}{\text{argmax}}\left \{ C_n^i \right \},i\in\left \{ 1,2,\cdots ,r \right \}
\end{equation} 
where $r$ is the number of classes. 

The CNN architecture allows us to have a more precise yet efficient classification algorithm using the extracted scalo-temporal features compared to raw time series data. The architecture of the CNN was designed to effectively capture both spatial and temporal features of the scalogram images. The network begins with an image input layer configured to handle $\zeta \times \zeta$ pixel images with ${c}'$ channels. This is followed by several convolutional layers, each equipped with ReLU activation functions and batch normalization to ensure stable and efficient training. Max-pooling layers are interspersed between the convolutional layers to progressively reduce the spatial dimensions of the feature maps while retaining the most salient features.

A crucial aspect of the developed architecture is the integration of a spatial attention mechanism. This mechanism is implemented using a combination of pooling and convolutional layers that learn to weight the importance of different parts of the input feature maps. These weights are used to modulate the feature maps, emphasizing the most informative regions and suppressing the less relevant ones. This attention-enhanced feature map is then multiplied by the original feature map, ensuring that the network focuses on the critical parts of the input data. Utilizing a CNN with attention mechanism allows for leveraging the strengths of both convolutional layers and attention mechanisms. Convolutional layers are proficient at capturing local spatial hierarchies in the input data, which is crucial for recognizing patterns in the scalogram images. However, they might not always capture the global context effectively. The attention mechanism compensates for this by enabling the model to weigh the importance of different spatial regions, thereby enhancing the network’s focus on the most relevant features. The developed architecture is beneficial due to its balanced approach to handling both local and global features of the input data, improving the network's ability to distinguish between different complexity levels in the input data. The developed CNN architecture with a spatial attention mechanism enhances the model's performance in classifying time series complexity levels, which is crucial for maintaining the security and integrity of NCSs. 

\begin{figure}[t]
 	\centerline{\includegraphics[width=8.9cm]{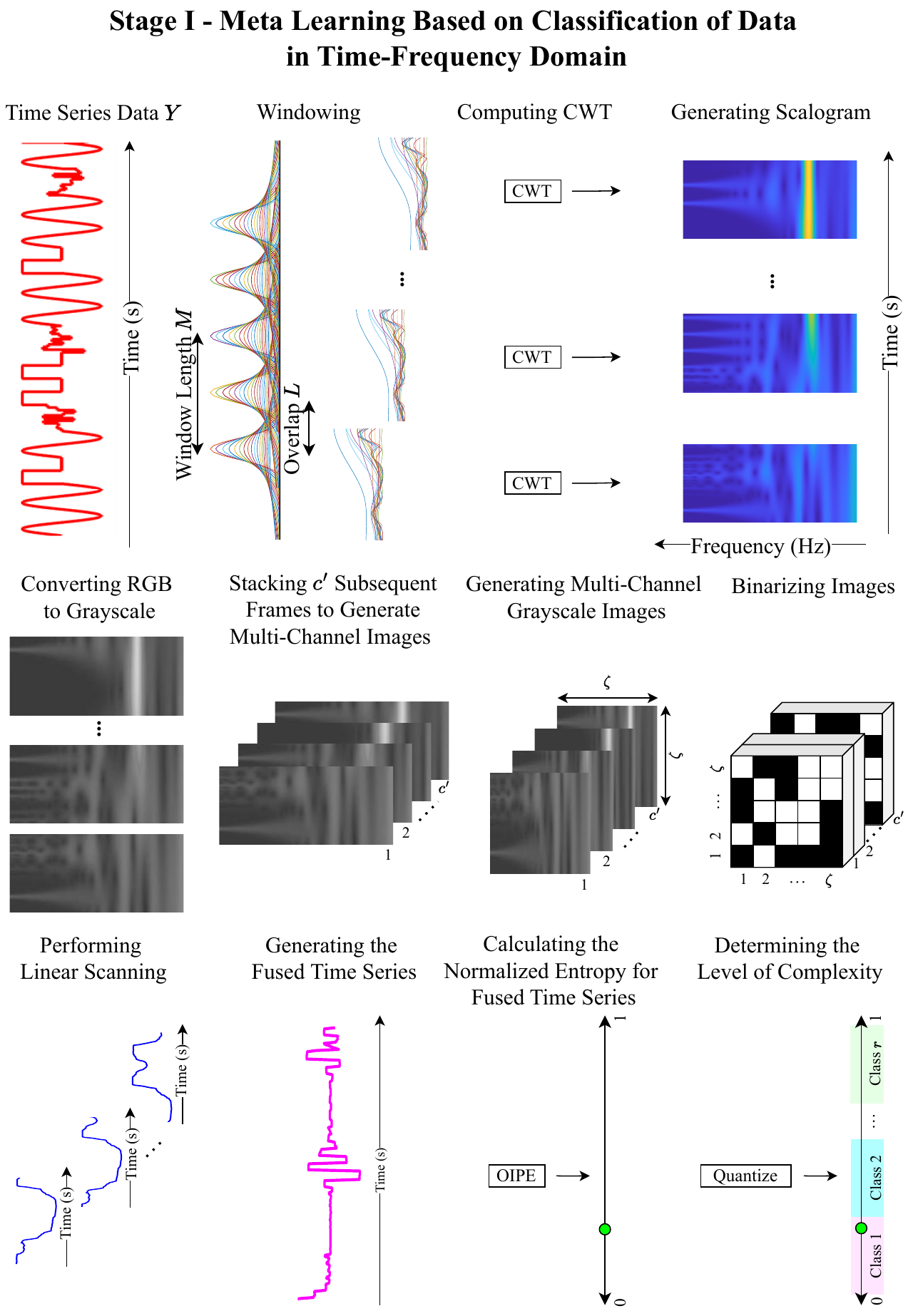}}
 	\caption{Stage I of proposed framework for online mitigation of FDI attacks.}
 	\label{fig32a}
\end{figure}

Given the detailed description of Stage I of the proposed framework, Fig. \ref{fig32a} summarizes this stage.

\section{Stage II - Mitigation of FDI Attacks} \label{sec3}
\subsection{Phase I: Design Procedure of Base Time Series Forecasting Models}
Mitigating FDI attacks requires the development of time series forecasting models that can accurately predict future steps in a sequence of data, leveraging the strengths of various deep learning architectures. By employing distinct time series forecasting architectures, ranging from a simple gated recurrent unit network with few layers to deep neural networks, the goal is to capture temporal dependencies within the data. The base time series forecasting models in the proposed stacked ensemble learning architecture are designed to enhance the diversity of the framework, addressing the varying complexity levels within the data. By incorporating multiple base models, each specialized for different level of complexity in data, the ensemble ensures that the predicted time series is highly accurate. At the same time, the structure of the ensemble is carefully developed to maintain computational efficiency, ensuring that the mitigation process is both effective and efficient, even in real-time applications.

\subsubsection{Gated Recurrent Unit (GRU) Network}
Recurrent Neural Networks (RNNs) are suitable for time series forecasting applications due to their ability to capture temporal dependencies among sequences of input signal. This is done through feedback connections between input units and hidden units, which allow the network to remember information from previous time steps and use it to process new input data. However, traditional RNNs may not perform well in handling complex time series data, and cannot be used for very long sequences of data due to vanishing gradient problem. To overcome these limitations, GRUs are introduced as modifications to traditional RNNs. These networks are able to model longer temporal dependencies in sequential data by selectively remembering and forgetting previous input data using two gates: the update gate and the reset gate. The update gate determines how much of the previous hidden state should be preserved, while the reset gate controls how much of the past information should be discarded. The update gate and reset gate of GRU networks perform the following calculations, respectively:
\begin{subequations} \label{eq3.4}
        \begin{align}
            z_t &= \sigma_s\left ( W_{zy}y_t + W_{zh}h_{t-1} + b_z \right ) \\
            r_t &= \sigma_s\left ( W_{ry}y_t + W_{rh}h_{t-1} + b_r \right )
        \end{align}
\end{subequations}
where $\sigma_s(\cdot)$ is the Sigmoid activation function, as defined in (\ref{eq3.5}), $ W_{zy}$ is the weight matrix that relates inputs to update gate, $W_{zh}$ relates hidden state to update gate, $b_z$ is the update gate bias term, $ W_{ry}$ relates inputs to reset gate, $W_{rh}$ relates hidden state to reset gate, and $b_r$ is the reset gate bias term. 
\begin{equation} \label{eq3.5}
    \sigma _s(x)=\frac{1}{1+e^{-x}}
\end{equation}
where $x$ is the input of the Sigmoid activation function. The hidden state $h_t$ is calculated based on the current input, the previous hidden state $h_{t-1}$, and the value of update and reset gates based on the following equation:
\begin{equation} \label{eq3.6}
    h_t=(1-z_t)\odot h_{t-1}+z_t\odot \tilde h_t
\end{equation}
where $\odot$ denotes the element-wise product, and $\tilde h_t$ is calculated using the following equation \cite{8053243}:
\begin{equation} \label{eq3.7}
     \tilde h_t=Tanh(W_{hy}y_t+W_{hh}(r_t\odot h_{t-1})+b_h) 
\end{equation}

Using the combination of update and reset gates allows GRUs to retain relevant information from long sequences and discard irrelevant or outdated information. The output at each time step $\hat y_t$ is calculated using the current hidden state $h_t$ based on the following equation:
\begin{equation} \label{eq3.3}
    \hat y_t=\sigma\left ( W_{\hat yh}h_t + b_{\hat y} \right )
\end{equation}
where $\sigma(\cdot)$ is an activation function, $W_{\hat yh}$ is the weight matrix that relates hidden units to the output, and $b_{\hat y}$ is the output bias term.

\subsubsection{LSTM Network} LSTM networks are a type of RNNs that use cell states, which function as the memory of the LSTM network, to store and output information. These cells are designed to remember information for long periods of time. LSTM networks consist of three gates: the input gate, forget gate, and output gate. These three gates provide a more flexible control on the flow of information compared to GRU structure, allowing the LSTM to remember or forget information over time using the following equations, respectively:
\begin{subequations} \label{eq3.8}
    \begin{align}
            i_t &= \sigma_s\left ( W_{iy}y_t + W_{ih}h_{t-1} + b_i \right ) \\
            f_t &= \sigma_s\left ( W_{fy}y_t + W_{fh}h_{t-1} + b_f \right ) \\ 
            o_t &= \sigma_s\left ( W_{oy}y_t + W_{oh}h_{t-1} + b_o \right ) 
\end{align}
\end{subequations}
where $\sigma_s(\cdot)$ is the Sigmoid activation function, $ W_{iy}$ is the weight matrix that relates inputs to input gate, $W_{ih}$ relates hidden state to input gate, $b_i$ is the input gate bias term, $ W_{fy}$ relates inputs to forget gate, $W_{fh}$ relates hidden state to forget gate, $b_f$ is the forget gate bias term, $ W_{oy}$ relates inputs to output gate, $W_{oh}$ relates hidden state to output gate, and $b_o$ is the output gate bias term. The cell state $c_t$ is calculated based on the current input, the previous hidden state $h_{t-1}$ and cell state $c_{t-1}$, and the value of input and forget gates based on the following equation:
\begin{equation} \label{eq3.9}
    c_t=f_t\odot c_{t-1}+i_t\odot Tanh(W_{cy}y_t+W_{ch}h_{t-1}+b_c) 
\end{equation}
where $\odot$ denotes the element-wise product, $ W_{cy}$ is the weight matrix that relates inputs to cell state, $W_{ch}$ relates hidden state to cell state, and $b_c$ is the cell state bias term. Then, the hidden state is calculated using the current cell state $c_t$ and the value of the output gate $o_t$ according to the following equation \cite{8053243}:
\begin{equation} \label{eq3.10}
    h_t=o_t\odot Tanh(c_t)
\end{equation}

The output at each time step $\hat y_t$ in LSTM is calculated similar to the GRU network, as defined in (\ref{eq3.3}). LSTM networks are proficient in learning complex relationships and capturing long-term dependencies in time series data. However, the larger number of parameters makes LSTMs slower during training and more prone to overfitting on small datasets. 

\subsubsection{CNN-LSTM Network}
In the LSTM and GRU networks discussed previously, time series data are directly fed into the network. These networks excel in learning temporal dependencies among sequences of input, but they are not suitable for capturing the frequency domain features as well. Succeeding a CNN with various recurrent layers, such as LSTM, addresses this issue through reducing the frequency variance in the input data by passing it through multiple convolutional layers. The convolutional layers act as feature extractors, identifying and enhancing important patterns in the input while reducing irrelevant frequency variations. Convolutional kernels are applied across different regions of the input to smooth out fluctuations and highlight consistent features, making time series data more stable for subsequent processing. The gradual abstraction provided by multiple convolutional layers allows the model to focus on meaningful sequential data characteristics, improving its ability to interpret complex time series data \cite{7178838}. After reducing the frequency variance in input data, the output of the CNN is then fed into LSTM layers, which are well-suited for capturing the temporal dependencies of the signal.

An important aspect of this network's design is the incorporation of residual connections. The CNN processes long-term features and creates a higher-order representation that is then passed into LSTM layers for time series forecasting, causing the short-term features to be ignored. To address this challenge, the short-term features are also directly passed into the LSTM by creating residual connections. By feeding both the original short-term features and the CNN-derived long-term features into the LSTM layer, the model's ability to handle more complex time series is enhanced \cite{7178838}.

\subsubsection{ResNet-50-LSTM Network}
Instead of the CNN network described in the previous part, deep residual networks can be leveraged for reducing the frequency variance and feature extraction from input data to predict time series data with high complexities. These networks incorporate skip connections, where the input from a previous layer is directly added to the output of the current layer. Using skip connections allows ResNets, or Residual Networks, for maintaining high performance while being more compact compared to traditional CNNs \cite{shafiq2022deep}. Moreover, the skip connections help address the problem of vanishing gradients in deep networks, allowing ResNets to be much deeper than traditional CNNs without suffering from performance degradation. Similar to the CNN-LSTM network, in this article the ResNet-50 network was succeeded with LSTM layers to perform time series forecasting. Increasing the number of layers in the network and developing more advanced network architectures increases the prediction capacity of the network and allows for extracting more complex features from input data. Consequently, this network is capable of predicting more complex time series data compared to the CNN-LSTM network. 

\subsubsection{Modified Xception-LSTM Network}
This article introduces a new network architecture inspired by Xception network, originally developed in \cite{8099678} for image classification tasks, for reducing the frequency variance and extracting features from input data. The Xception network uses the concept of depth-wise separable convolutions, which is a depth-wise convolution followed by a point-wise convolution. One can refer to \cite{8099678} for a detailed description and design procedure of this network. The Xception architecture is modified by removing repeated middle flow blocks to reduce the computational complexity and training time, and changing output layers to adapt it for time series forecasting. By doing this, the inference time is also significantly reduced which is crucial for real-time prediction of time series data. Similar to the previous network, LSTM layers are used to adapt the modified Xception network for time series forecasting.

\subsection{Phase II: Model Assignment to Distinct Complexity Levels}
In the proposed stacked ensemble learning architecture, model assignment refers to the offline procedure of evaluating the performance of different time series forecasting models on each level of complexity of the data, which is determined by the complexity classifier CNN network. Once the base time series forecasting models are designed and trained, model assignment is performed offline by evaluating the performance of base models in time series forecasting on the validation subset, and measuring the Root Mean Squared Error (RMSE) metric for reconstruction of each level of time series data complexity, where the RMSE between two vectors $z$ and $\check{z}$ with length $N_z$ is expressed by the following equation:
\begin{equation} \label{eq3.11}
    \text{RMSE}(z,\check{z})=\sqrt{\frac{1}{N_{z}}\sum_{j=1}^{N_{z}}\left ( z_j-\check{z}_j \right )^{2}}
\end{equation}
Another factor that needs to be considered for model assignment is the computational burden of using a base model for prediction. Using a more intricate base model provides greater prediction capacity at the expense of more computational burden. Consequently, model assignment is performed by taking both the RMSE metric and average inference time of base models into consideration. The base model that achieves the optimal balance between performance and speed for each complexity level is selected to predict time series data corresponding to that level of complexity. This model is referred to as the optimal predictor, denoted by $P^{opt}_i$ for $i^{\text{th}}$ level of complexity, and is determined as follows:
\begin{equation} \label{eq3.12}
    P^{opt}_i = \underset{l}{\text{argmin}}\left \{ \varepsilon \times \text{RMSE}(a^{v_i},\hat a^{l})+ (1-\varepsilon) \times T^l \right \}
\end{equation}
where $\varepsilon$ is a coefficient which determines the balance between performance and speed of the framework, $a^{v_i}$ is the vector of data points in the validation dataset belonging to $i^{\text{th}}$ complexity level, $\hat{a}^l$ is the predicted value using $l^{\text{th}}$ base model, $T^l$ is the average inference time of $l^{\text{th}}$ base model, $l\in\left \{ 1,2,\cdots ,n \right \}$, and $i\in\left \{ 1,2,\cdots ,r \right \}$. Furthermore, threshold values $\tau_{c}^i$ need to be determined offline for model selection. These threshold values represent the minimum required normalized confidence score that a data point belongs to the $i^{\text{th}}$ class. The value of these thresholds is determined by evaluating the classification accuracy of each level of time series data complexity on the validation dataset. 

\subsubsection{Relation Between Complexity Levels and Base Model Architectures}
The developed stacked ensemble learning architecture consists of a diverse range of base time series forecasting models, varying from a simple GRU network with only few layers to intricate neural network structures. The following factors contribute to allowing a base model for predicting more complex time series data: (i) incorporating recurrent layers with more advanced structures, allowing for a more flexible control on the flow of information and handling more complex data, (ii) increasing the number of layers within a network, leading to a greater prediction capacity, (iii) including convolutional layers, leading to reduction of frequency variance in input data and allowing the model to focus on meaningful sequential data characteristics, and (iv) incorporation of residual connections, allowing for learning both short-term and long-term features in input data.

Table \ref{tab3.1} provides a comparison of the base time series forecasting models in terms of the above design factors to demonstrate their effectiveness in handing complex time series data.
\begin{table}[t]
	\caption{Comparison of base models in terms of design factors contributing to their effectiveness in handling complex data}
	\begin{center}
	\label{tab3.1}
\begin{tabular}{|l|c|c|c|c|}
	\hline
	 \multicolumn{1}{|c|}{Design Factor} & (i) & (ii) & (iii) & (iv)\\
	\hline
      GRU Network & \ding{55} & \ding{55} & \ding{55} & \ding{55} \\

      LSTM Network & \ding{51} & \ding{55} & \ding{55} & \ding{55} \\
 
      CNN-LSTM Network & \ding{51} & \ding{55} & \ding{51} & \ding{51} \\
      
      ResNet-50-LSTM Network & \ding{51} & \ding{51} & \ding{51} & \ding{51} \\
      
      Modified Xception-LSTM Network & \ding{51} & \ding{51} & \ding{51} & \ding{51} \\
	\hline
\end{tabular}
\end{center}
\end{table}

\subsection{Phase III: Online Time Series Forecasting}
\subsubsection{Model Selection}
In contrast to model assignment, model selection is the online process where, for incoming data, the appropriate model is chosen dynamically based on the complexity level of the input data. This allows for real-time selection of the most suitable time series forecasting model from the stack for accurate and efficient forecasting. To perform model selection, for each window of the time series data, it is converted to a multi-channel scalogram image as defined in (\ref{eq15}). Then, it is fed to the developed complexity classifier CNN network to compute the normalized confidence scores defined in (\ref{eq33}). The model selection is performed as follows:
\begin{equation} \label{eq3.13}
    P^{opt}=\left\{
\begin{array}{ll}
 P^{opt}_i, &  C_n^i \geqslant  \tau_{c}^i \\ 
\frac{1}{n}\sum_{i=1}^{n}P^{opt}_i, &  C_n^i < \tau_{c}^i
\end{array}
\right.
\end{equation}
where $\tau_c^i, \ i \in \left\{ 1,2,\cdots , r \right\}$ represents a set of threshold values applied to the normalized confidence scores. These threshold values are chosen by running the trained CNN model on the validation dataset.

\subsubsection{Time Series Forecasting}
\begin{figure}[t]
 	\centerline{\includegraphics[width=8.9cm]{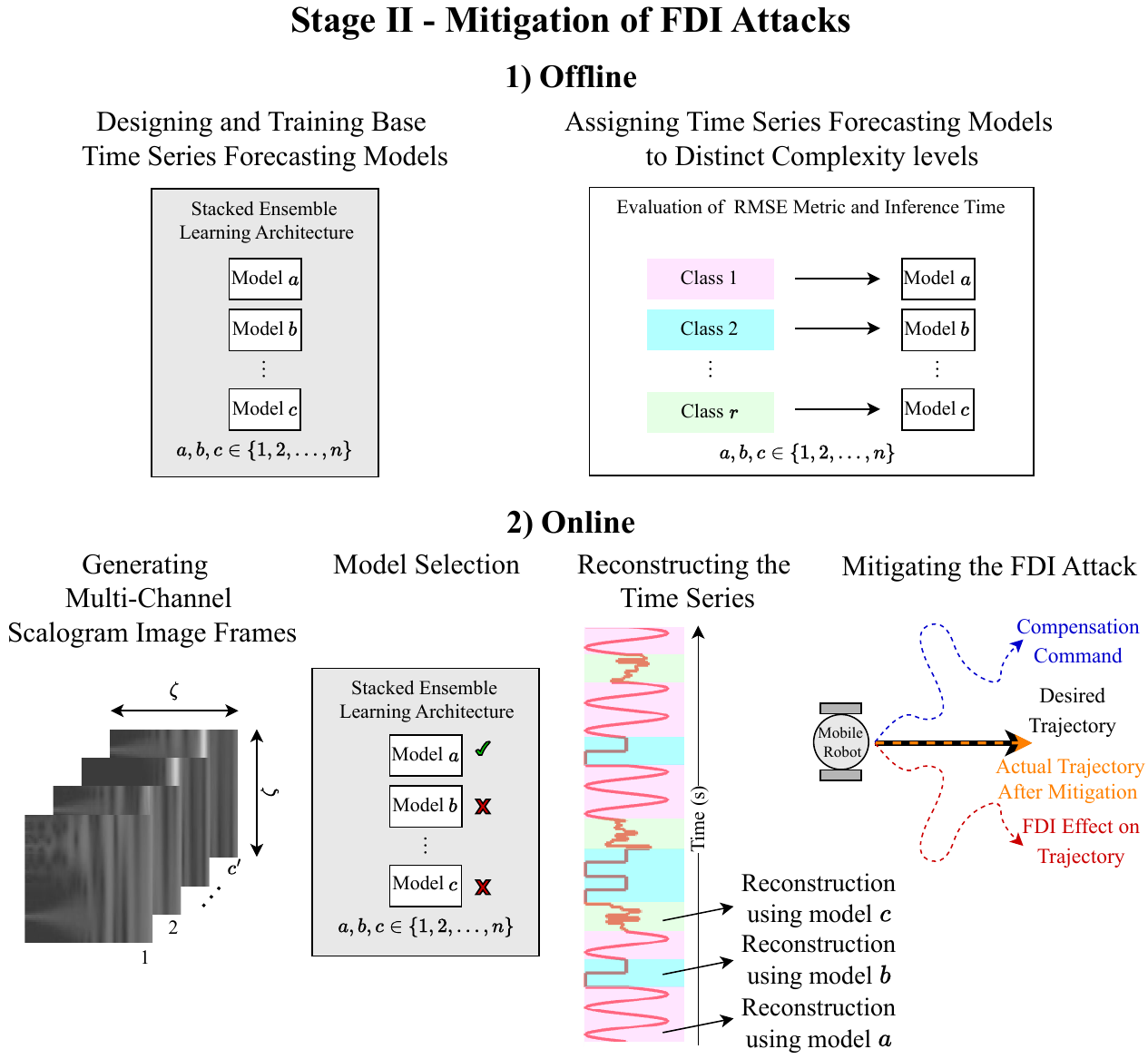}}
 	\caption{Stage II of proposed framework for online mitigation of FDI attacks.}
 	\label{fig32}
\end{figure}

Having the time series data $Y$, as defined in (\ref{eq2.1}), the goal is to predict $q$-steps ahead of the sequence. The scalo-temporal features are extracted from the time series data, providing a comprehensive understanding of the underlying patterns and dependencies. Once these features are identified, the model selection is performed to determine the most suitable time series forecasting model. This selected model is then employed to predict the future values of the time series, i.e. performing a $q$-steps ahead prediction, which is represented as $\hat Y^{(q)}= \left \{\hat y_{t+q}, t\in I \right \}$. To address and counteract the effects of FDI attacks, a correction mechanism can be incorporated into the proposed framework. This mechanism functions by detecting deviations in the predicted values, which are indicative of potential FDI attack occurrences. By comparing the predicted values $\hat Y^{(q)}$ with the actual measurements $Y$, the system is able to reconstruct the FDI attack using the following equation:
\begin{equation} \label{eq3.15}
    \hat A = \left\{ \hat a_t,\ t \in I \right\},\quad \hat a_t = \tilde y_t - \hat y_t
\end{equation}
Then, the system applies the following correction equation to mitigate the impact of FDI attacks, ensuring the integrity of the generated control commands:
\begin{equation} \label{eq3.16}
     \breve Y = \left\{ \breve y_t,\ t \in I \right\},\quad \breve y_t = \tilde y_t - \hat a_t
\end{equation}
This ensures that the system's operation remains safe despite the presence of FDI attacks. Algorithm \ref{alg3} summarizes Stage II of the proposed framework, and detailed description of this stage is depicted in Fig. \ref{fig32}.  

\begin{algorithm}[t]
\caption{Mitigation of FDI Attacks}\label{alg3}
\textbf{Input:} $Y$ \\ \hspace{-0.11 cm} 
\textbf{Output:} $\breve Y$ \\
\textsc{\textbf{PHASE I:} Design Procedure of Base Time Series Forecasting Models}
\begin{algorithmic}[1]
    \STATE Define the length of time series $N_y$, the number of time series forecasting models $n$, prediction horizon $q$, and the number of complexity levels $r$
    \STATE Split the dataset into training and validation
    \STATE Train individual time series forecasting models
 
    \hspace{-0.7cm} \textsc{\textbf{PHASE II:} Model Assignment to Distinct Complexity Levels}
    \STATE Define coefficient $\varepsilon$
    \FOR{$ ( 1\le i \le r) $}
        \STATE Determine the optimal predictor $ P^{opt}_i$ based on the validation data using (\ref{eq3.12})
        \STATE Determine the normalized confidence score threshold $\tau_c^i$ based on validation data  
    \ENDFOR
    
    \hspace{-0.7cm} \textsc{\textbf{PHASE III:}  Online Time Series Forecasting}
    \STATE Define the minimum normalized confidence scores $\tau_c^i$
    \FOR{$ ( 1\le t \le N_y) $}
    \STATE Generate multi-channel scalogram image frame using (\ref{eq4})-(\ref{eq15})
    \STATE Calculate the normalized confidence scores using (\ref{eq33}) 
    \STATE Perform model selection using (\ref{eq3.13}) to determine $P^{opt}$
    \STATE Compute the prediction of time series $\hat y_{t+q}$ using the selected model $P^{opt}$
    \STATE Calculate the reconstructed FDI $\hat a_t$ using (\ref{eq3.15})
    \STATE Generate the corrected time series $\breve y_t$ using (\ref{eq3.16})
    \ENDFOR
\end{algorithmic}
\end{algorithm}

\section{Results and Discussions} \label{sec4}
In this section, the effectiveness of the proposed framework is demonstrated via numerical simulations in Simulink environment of MATLAB software. A leader-follower formation of differential drive mobile robots is used as the case study in this paper. A group of four mobile robots, including a leader and three follower robots aim to reach a predefined destination from their starting point. The mobile robots communicate their position via a network to maintain a certain formation during their movement. The communication channels enable these agents to perform a coordinated task efficiently. However, this makes them vulnerable to various types of cyberattacks. The differential drive mobile robots follow the nonlinear model below:
\begin{equation} \label{eq301}
    \begin{bmatrix}
x_{1,t+1}\\ 
x_{2,t+1}\\ 
x_{3,t+1}
\end{bmatrix}=\begin{bmatrix}
x_{1,t}+\nu_{t}cos(x_{3,t+1})\times dt\\ 
x_{2,t}+\nu_{t}sin(x_{3,t+1})\times dt\\ 
x_{3,t}+\omega_t \times dt
\end{bmatrix}
\end{equation}
where $x_{1,t}, x_{2,t}, x_{3,t}, \nu_t$, and $\omega_t$ are the position along X and Y axis, orientation, linear and angular velocity of the mobile robot, respectively \cite{HOSSEINZADEHYAMCHI2017194}. To control the mobile robots, it is desirable to express the equations of motion in terms of linear and angular velocity of the wheels. The linear and angular velocity of the mobile robot are related to the linear and angular velocity of the left and right wheels using the following equation:
\begin{equation} \label{eq302}
\left\{\begin{aligned}
\nu_t&=\frac{R}{2}(\omega_{R,t}+\omega_{L,t})\\ 
\omega_t &=\frac{\nu_t}{R}=\frac{R}{O}(\nu_{R,t}-\nu_{L,t})
\end{aligned}
\right.
\end{equation}
where $R,O,\omega_{R,t}, \omega_{L,t}, v_{R,t}$, and $v_{L,t}$ are the robot wheel radius, wheelbase, angular and linear velocity of right and left wheels, respectively. The mobile robot is controlled by adjusting the following inputs:
\begin{equation} \label{eq4.18}
    u_t=\begin{bmatrix}
u_{1,t} \\
u_{2,t}
\end{bmatrix}=\left\{\begin{aligned}
    \omega_{L,t}&=\frac{1}{R}(\nu_t-\frac{\omega_t O}{2})\\ 
    \omega_{R,t} &=\frac{1}{R}(\nu_t+\frac{\omega_t O}{2})
    \end{aligned}
    \right.
\end{equation}

The numerical values for the parameters of the differential drive mobile robot used in the simulations are summarized in Table \ref{tab4.4}. These parameters were selected to represent realistic conditions, forming the basis for the analysis and results discussed in this paper.
\begin{table}[t]
	\caption{Mobile robot parameters}
	\begin{center}
	\label{tab4.4}
\begin{tabular}{|c|c|c|}
	\hline
	 Parameter & Description & Value \\
	\hline
	 $R$ & Robot wheel radius & 10 cm \\
	
	 $O$ & Robot Wheelbase & 50 cm \\
	
	 $T_\text{s}$ & Sampling period & 50 ms \\
	
	 $N$ & Number of robots & 4 \\
	
	 $(x_\text{1,init},x_\text{2,init})$ & Initial position  & (40,45) \\
	
	 $(x_\text{1,d},x_\text{2,d})$ & Destination  & (4,7) \\
	\hline
\end{tabular}
\end{center}
\end{table}

The control objective of each mobile robot is following the position of leader robot with a specific distance, where the reference signal $v^i_t$ is generated by $\mathcal{DC}^i$ according to the following equation:
\begin{equation} \label{eq4.12}
        v_{t+1}^{i} = v_{t}^{i}+\psi_{t}^i 
\end{equation}
where $\psi_{t}^i$ is determined using the communicated variables from neighborhood, based on the following equation \cite{9550364}:
\begin{equation} \label{eq4.13}
    \psi_{t}^{i}=\sum_{j \in \mathcal{N}_i} \alpha_{ij}(y_t^{[j,i]}-y_t^{i})+b_t^i
\end{equation}
where $\mathcal{A}=\left [ \alpha_{ij} \right ]_{N\times N}$ is referred to as the adjacency matrix and determines the topology of the network, and $\mathcal{B}_t=[b_t^1, b_t^2, ..., b_t^N]^T$ is a bias term and can be adjusted to achieve different formations of mobile robots over time. The adjacency matrix is defined using the following equation:
\begin{equation} \label{eq4.19}
    \mathcal{A}=[\alpha_{ij}]_{N \times N}=\left\{
\begin{array}{ll}
 1, &  j \in \mathcal{N}_i \\ 
0, &  j \notin \mathcal{N}_i
\end{array}
\right.
\end{equation}

In this paper, a leader-follower formation control problem with $N=4$ robots is used as the case study with the following topology:
\begin{equation} \label{eq4.19}
\mathcal{A}=\begin{bmatrix}
0 & 0 & 0 & 0 \\
1 & 0 & 1 & 1 \\
1 & 1 & 0 & 1 \\
1 & 1 & 1 & 0 \\
\end{bmatrix},\mathcal{B}_t=\begin{bmatrix}
0 & 0 \\
3 & 0 \\
3 & 3 \\
0 & 3 \\
\end{bmatrix}
\end{equation}

The two vulnerabilities that could compromise the system's security are incorporated in (\ref{eq4.13}), and defined as follows:
\begin{enumerate}
    \item \textit{Scenario I: The communication link between the sensor and controller unit:} It is assumed that the communication link between the sensor and controller unit within subsystems is exposed to attackers, and they manage to inject signal $ a_t^{i} $ into communicated variables. Therefore, the following compromised variables are received by $\mathcal{DC}^i$:
    \begin{equation} \label{eq4.6}
    	\tilde y^i_t = y_t^{i} + a_t^{i}
    \end{equation}
    
    \item \textit{Scenario II: The communication link between neighboring subsystems:} Assuming that the communication link between neighboring subsystems is exposed to attackers, and they manage to inject signal $ a_t^{j} $ into communicated variables, the following compromised variables are received by the $i^{th}$ subsystem:
    \begin{equation} \label{eq4.7}
    	y_t^{[j,i]} = y_t^{j} + a_t^{j}\ ,\quad \tilde y^j_t=y_t^{[j,i]}
    \end{equation}
     where $ y_t^{[j,i]}, j \in \mathcal{N}_i$ is the output received from $j^{\text{th}}$ subsystem at $\mathcal{S}^i$.
\end{enumerate}

\begin{figure}[t]
  \begin{center}
    \includegraphics[width=5cm]{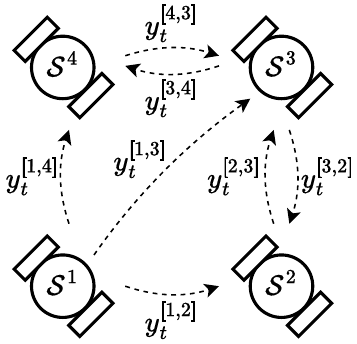}
    \caption{\label{fig23b} The topology of the mobile robots.}
  \end{center}
\end{figure}

The formation control problem studied in this paper along with the topology of the mobile robots is depicted in Fig. \ref{fig23b}.
 
\subsection{Stage I - Meta Learning Based on Classification of Data in Time-Frequency Domain}
\subsubsection{Phase I: Scalo-Temporal Feature Extraction using Continuous Wavelet Transform}
\begin{figure}[t]
	\centerline{\includegraphics[width=9cm]{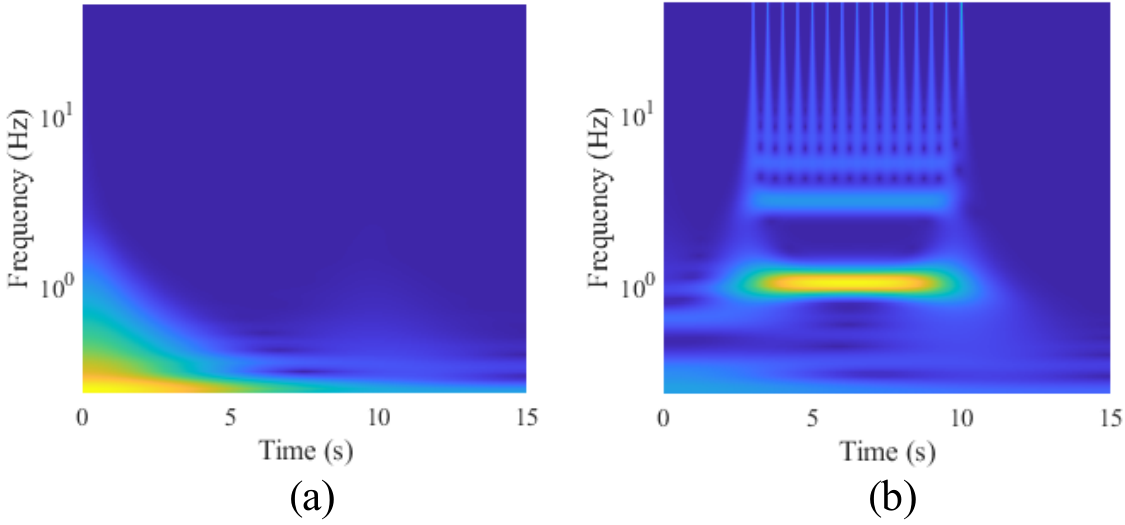}}
	\caption{Scalogram generated for: (a) Normal condition, (b) Under FDI attack.}
	\label{fig33}
\end{figure}

In order to perform meta learning, a dataset containing $N_y=15000$ samples from the sensor measurements of the mobile robot position along X axis with sampling period $T_s=10\ ms$ was generated. This time series data was split into $r=3$ equal partitions, each consisting of 5000 samples that included distinct complexity levels, represented by a sum of sinusoidal waves, a sum of square waves, and a white noise signal. The frequency of the signals in all three partitions varied randomly within the 0.5-2 Hertz range, and the amplitude was randomly selected between 0.5 and 10 meters. A CWT filter bank was created and the windowing operation on the time series data was performed using a window length of $M=60$ and a large overlap of $L=54$ samples to increase the resolution in time domain, and created $N_m=2492$ frames of data. Then, the Morse wavelet with $S=92$ scales, as expressed in (\ref{eq8}), was applied on each window and the scalogram frames of data, $SC^{(i)} \in \mathbb{R}^{92\times 60}, i\in \left \{ 1,2,\cdots ,2492 \right \}$, were generated using (\ref{eq6})-(\ref{eq9}). Fig. \ref{fig33} depicts the generated scalogram for data in normal condition and in the presence of FDI attack.

To prepare the data for training the CNN FDI complexity classifier, the image frames were converted to grayscale using (\ref{eq12}) to reduce the number of channels from three to one. To incorporate the dynamic spectrum of the data and increase the robustness of the classification algorithm, ${c}'=8$ frames were stacked to form the multi-channel scalogram image frames. Then, these image frames were resized with $\zeta$ chosen as 227 using (\ref{eq14}). The generated dataset was then partitioned into training (70\%) and validation (30\%) subsets. 

\subsubsection{Phase II: Complexity Analysis using Entropy Metric}
To obtain the corresponding complexity level \(L_c^{(i)}\) for each multi-channel scalogram image frame \(P_g^{(i)}\), as outlined in Algorithm \ref{alg5}, the frames were initially binarized using the threshold \(\tau_b = 128\). The resulting binary image frames \(P_b^{(i)}\) were then used to create the fused time series \(Y_f^{(i)}\), as detailed in (\ref{eq17})-(\ref{eq19}). The OIPE algorithm was applied to the fused time series, following (\ref{eq20})-(\ref{eq28}), to analyze the complexity of the fused data. This analysis used an embedding dimension \(D = 4\), a time delay \(\tau = 1\), and \(H = 32\), with the number of quantization levels selected through a \(K = 10\) fold cross-validation. Then, the calculated entropy values were quantized to generate the $r=3$ complexity levels $L_c^{(i)}$. This step was done using the validation dataset, with thresholds set at $\tau_e^1=0.678, \tau_e^2=0.892$. Consequently, as outlined in (\ref{eq28}), a level 1 complex time series is characterized by an entropy below 0.678, a level 2 complex time series has an entropy ranging from 0.678 to less than 0.892, and a level 3 complex time series possesses an entropy between 0.892 and 1. 

\subsubsection{Phase III: FDI Complexity Classification using CNN}
Following the steps described in the previous phases,  $N_t= \frac{10500-54}{60-54}=1741$ image frames and their corresponding complexity levels were generated and given as input to the designed complexity classifier CNN model, the layer types and output sizes of which are summarized in Table \ref{tab4.1}.
\begin{table}[t]
    \caption{Layer types and output sizes of the complexity classifier CNN model}
    \begin{center}
    \label{tab4.1}
    \begin{tabular}{cc}
    \hline
    Layer Type                   & Output Size                \\ \hline
    Input                        & (227, 227, 8)              \\ \hline
    Conv-2D                      & (227, 227, 64)             \\
    Normalization                & (227, 227, 64)             \\
    Spatial Attention            & (227, 227, 64)             \\
    Max Pooling                  & (113, 113, 64)             \\ \hline
    Conv-2D                      & (113, 113, 64)             \\
    Normalization                & (113, 113, 64)             \\
    Spatial Attention            & (113, 113, 64)             \\
    Max Pooling                  & (56, 56, 64)               \\ \hline
    Conv-2D                      & (56, 56, 128)              \\
    Normalization                & (56, 56, 128)              \\
    Spatial Attention            & (56, 56, 128)            \\ \hline
    Flatten                      & (1,634,432)                \\ \hline
    Fully Connected              & (3, 1, 1)                  \\ \hline
    Output                       & (3, 1, 1)                  \\ \hline
    \multicolumn{2}{c}{Total Trainable Parameters: 1,320,195} \\ \hline
    \end{tabular}
    \end{center}
\end{table}

To evaluate the performance of the complexity classifier CNN model, a test dataset containing time series data with switching complexity over time was generated. The dataset consists of $1501$ samples, where samples 201-300, 501-700, and 1001-1100 include a level 1 complex time series, a signal with level 2 complexity is added to samples 401-500, 701-800 and 1301-1400, and samples 1-200, 301-400, 801-1000, 1101-1300, and 1401-1500 are contaminated with level 3 complex signal to create the time series data with switching complexity level. Due to the presence of multiple transitions between the complexity levels, a robust classification algorithm is needed to distinguish between distinct complexity levels. For this purpose, Algorithm \ref{alg4} was applied to the test data and $N_I=241$ multi-channel scalogram image frames were generated to be fed to the complexity classifier CNN model. The accuracy of the classification algorithm is defined as follows:
\begin{equation} \label{eq402}
    \text{Accuracy}=\frac{N_C}{N_I} \times 100
\end{equation}
where $N_C$ is the total number of correctly classified frames by the CNN model. The proposed complexity classifier described in Table \ref{tab4.1} showed an accuracy of 99.53\% on the validation data, and 96.68\% on the test data, which demonstrates its effectiveness in classifying distinct complexity levels. A more powerful tool in analysing the classification performance is using a confusion matrix as described in Table \ref{tab4.2}, where Pred. refers to the predicted label, and Precision and Recall for class $a$ in a multi-class classification problem are defined as follows:
\begin{table}[t]
	\caption{Confusion matrix for the complexity classifier CNN model}
	\begin{center}
	\label{tab4.2}
\begin{tabular}{|cc|ccc|c|}
\hline
\multicolumn{2}{|c|}{\multirow{2}{*}{Complexity Classifier CNN}} & \multicolumn{3}{|c|}{Actual}    & \multirow{2}{*}{Precision} \\
\multicolumn{2}{|c|}{}                                        & class 1 & class 2 & class 3 &                            \\ \hline
\multirow{3}{*}{\rotatebox{90}{Pred.}}           & Class 1            & 76         & 7   & 1           & 90.5                          \\
                                     & Class 2                   & 0          & 57  & 0           & 100                      \\
                                     & Class 3           & 0          & 0   & 100          & 100                          \\ \hline
\multicolumn{2}{|c|}{Recall}                                  & 100      & 89.1   & 99.1       &                            \\ \hline
\end{tabular}
\end{center}
\end{table}

\begin{equation} \label{eq404}
\begin{matrix}
\text{Precision}_{\text{class}\ a} = \frac{TP_{\text{class}\ a}}{TP_{\text{class}\ a}+FP_{\text{class}\ a}} \times 100 \\ \\
\text{Recall}_{\text{class}\ a} = \frac{TP_{\text{class}\ a}}{TP_{\text{class}\ a}+FN_{\text{class}\ a}} \times 100
\end{matrix}
\end{equation}
where $TP, FP$, and $FN$ refer to the number of True Positive, False Positive, and False Negative prediction results, respectively. According to Table \ref{tab4.2}, the proposed framework is able to predict the complexity level of time series data precisely and can be used to dynamically select the base time series forecasting model for FDI mitigation.

\subsection{Stage II - Mitigation of FDI Attacks} 
\subsubsection{Phase I: Design Procedure of Base Time Series Forecasting Models}
\begin{figure}[t]
 	\centerline{\includegraphics[width=8.8cm]{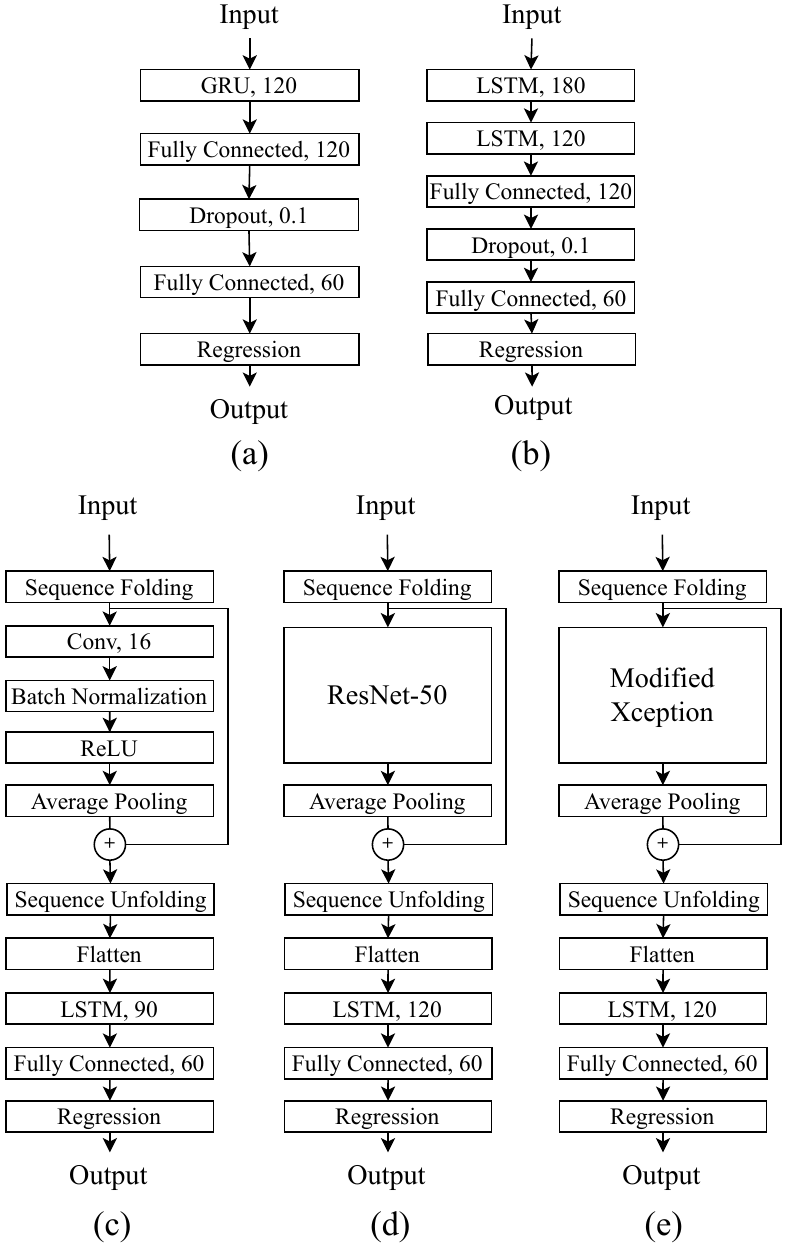}}
 	\caption{Designed models for stacked ensemble learning architecture: (a) GRU, (b) LSTM, (c) CNN-LSTM, (d) ResNet-50-LSTM, (e) Modified Xception-LSTM.}
 	\label{fig2}
 \end{figure}

To implement the proposed framework, the first step is developing the stacked ensemble learning architecture. An ensemble of five different network architectures was developed for this purpose, that range from a single-layer LSTM network to deep ResNet-50 network as a backbone for feature extraction and a recurrent structure for time series forecasting. To train the base models, a the training subset containing 10500 samples was used. As mentioned in section \ref{sec3}, each of base models takes a sequence of position sensor measurements of robot as input, and then predicts the FDI attack using a $q$-steps ahead time series forecasting. In the simulations, the sensor measurements from last 60 samples were used for time series forecasting, and one step ahead of the data was predicted, leading to $ k =60 $ and $ q=1 $. Fig. \ref{fig2} depicts the detailed structure of the base models embedded in our proposed stacked ensemble learning architecture.

\subsubsection{Phase II: Model Assignment to Distinct Complexity Levels}
To perform model assignment, the validation data were divided into three subsets based on their complexity levels. Then, the RMSE between the actual and predicted time series of each complexity level as well as the average inference time for each base model were calculated, and the base model achieving the optimal balance between performance and speed using $\varepsilon=0.4$ was selected as the optimal predictor $P^{opt}_{i}$, as described in (\ref{eq3.12}). The models assigned for prediction of each complexity level, along with their corresponding RMSE values and average inference times, are summarized in Table \ref{tab4.3}, where $T_{p}$ denotes the average inference time for performing time series forecasting. With these model assignments, the proposed framework can dynamically select the base time series forecasting model for FDI mitigation based on the complexity of the input data in real-time. Moreover, if the complexity classifier CNN model fails to predict the complexity level of the input data with a confidence score more than $\tau_c^i=0.7,\ i\in\left \{ 1,2,3 \right \}$, the average of the predictions from all networks in the stacked architecture is used for FDI mitigation, as described in (\ref{eq3.13}).
\begin{table}[t]
	\caption{Model assignment using $\varepsilon=0.4$}
	\begin{center}
	\label{tab4.3}
    \begin{tabular}{|c|c|c|c|}
    	\hline
    	 Complexity Level & Assigned Model & RMSE & $T_{p}$\\
    	\hline
    	 Class 1 & Modified Xception-LSTM & 0.301 m & 10.44 ms\\
    	
    	 Class 2 & LSTM & 0.355 m & 5.13 ms\\
    	
    	 Class 3 & GRU & 0.171 m & 4.93 ms\\
    	\hline
    \end{tabular}
    \end{center}
\end{table}
\subsubsection{Phase III: Online Time Series Forecasting}

\textit{\\ 3.1) Scenario I:} In this part, the case where the leader robot, i.e., $\mathcal{S}^1$, is the target of FDI attack on its position sensor measurements is considered. Targeting the leader of a formation can pose a serious threat against the safety of all of robots, as it can lead to deviation of all the robots in the formation from their desired trajectory, and preventing them to reach their destination. Furthermore, the leader robot cannot simply be isolated from the network and enforced to shutdown because other robots are receiving their reference trajectory from the leader. Therefore, it is essential to ensure that the leader is equipped with resilient control algorithms and can maintain normal operation even in the presence of FDI attacks.

\begin{figure}[t]
 	\centerline{\includegraphics[width=8.8cm]{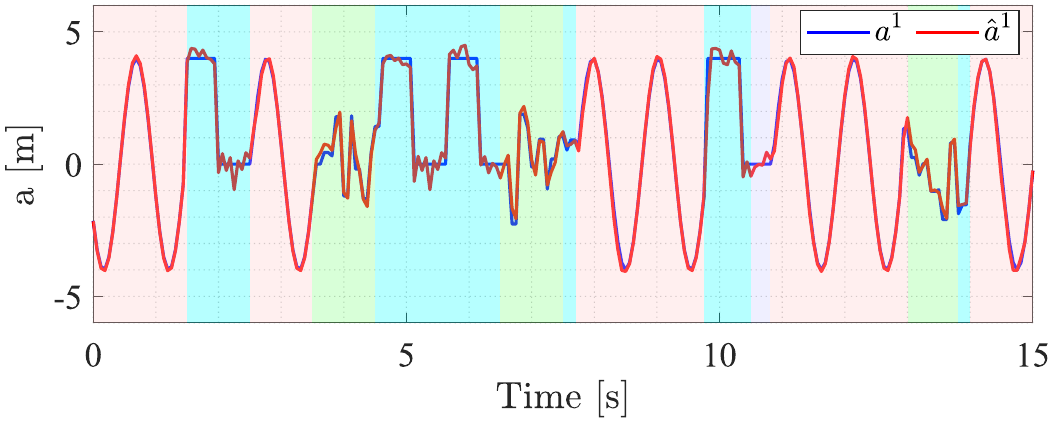}}
 	\caption{Reconstruction of the FDI attack by dynamically selecting the time series forecasting model.}
 	\label{fig100}
 \end{figure}

To demonstrate the effectiveness of the proposed framework, a comparison was performed between the FDI attack from the test dataset and the predicted FDI attack utilizing the assigned models, as depicted in Fig. \ref{fig100}, where the segments with blue, green, and red background correspond to the predicted complexity levels $\hat{L}_c$ using the complexity classifier CNN model, representing levels 1, 2, and 3, respectively. The prediction performance was evaluated through the RMSE metric between the actual and predicted FDI attack, and the simulation results are summarized in Table \ref{tab4.5}, where the improvement percentage obtained by utilizing the proposed framework in comparison with the individual base models, denoted by Imp, is calculated by the following equation:
\begin{equation} \label{eq4.10}
    \text{Imp} = \frac{\left| \text{RMSE}(a,\hat a^{\text{stacked}})-\text{RMSE}(a,\hat a^{l})\right|}{\text{RMSE}(a,\hat a^{l})}\times 100
\end{equation}
where $\hat a^{\text{stacked}}$ is the predicted FDI attack using the proposed stacked ensemble learning architecture, $l\in\left \{ 1,2,\cdots ,n \right \}$, and $| \cdot |$ denotes the absolute value.
 
 \begin{table}[t]
	\caption{Comparison between RMSE metric on test dataset using individual base models and the proposed framework}
	\begin{center}
	\label{tab4.5}
\begin{tabular}{|l|c|c|}
	\hline
	\multicolumn{1}{|c|}{Model} & RMSE & Imp\\
	\hline
      \hspace{0pt}Stacked Architecture & 0.287 m & -\\
      
      GRU & 0.377 m & 23.87\% \\
      
      LSTM & 0.323 m & 11.15\% \\
		
      CNN-LSTM & 0.576 m & 50.17\% \\

      ResNet-50-LSTM & 0.407 m & 29.49\% \\

      Modified Xception-LSTM & 0.414 m & 30.68\% \\    
      
	\hline
\end{tabular}
\end{center}
\end{table}

\begin{figure}[t]
	\centerline{\includegraphics[width=8.8cm]{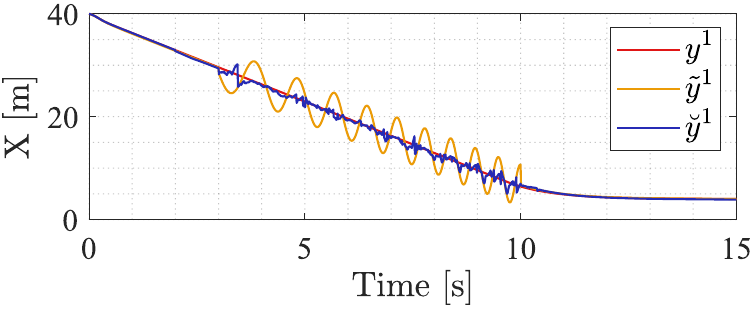}}
	\caption{Comparison of trajectory of the robot using proposed framework with reference and uncompensated trajectories.}
	\label{fig6}
\end{figure}
As mentioned previously, the proposed two-stage framework aims to mitigate the FDI attacks on the position sensor measurements of the mobile robots. To investigate the effectiveness of the proposed framework in real-time mitigation of FDI attacks, a chirp signal with amplitude 4 varying from 0.2 to 2 Hertz in a 7 seconds interval was injected to data of position of $\mathcal{S}^1$ along the X axis. The simulation was run for 15 seconds, where the FDI attack started at $ t = 3\ s $ and ended at $ t = 10\ s $. Fig. \ref{fig6} presents a comparison of the reference trajectory, the effect of FDI attack on position sensors of the robot, and its trajectory after the proposed framework is applied to it. As we can see in Fig. \ref{fig6}, in spite of presence of a malicious FDI attack with a relatively large amplitude on position sensor measurements, the robot stays close to its desired trajectory, which demonstrates the effectiveness of the proposed framework in mitigating the FDI attacks in real-time. 

\begin{figure}[t]
	\centerline{\includegraphics[width=8.8cm]{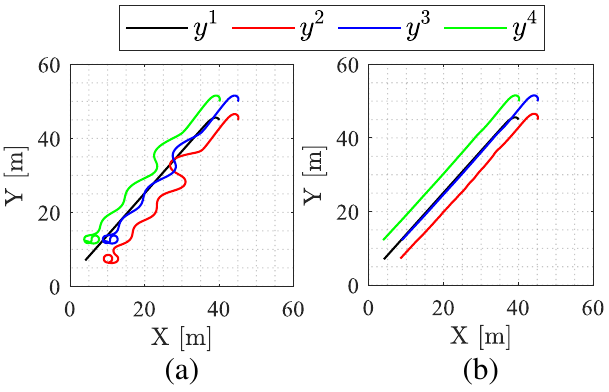}}
	\caption{Comparison between the trajectory of mobile robots: (a) No FDI mitigation, (b) FDI mitigation using the proposed framework.}
	\label{fig81}
\end{figure}

\textit{\\ 3.2) Scenario II:}
To demonstrate the effectiveness of the proposed framework, the performance of the framework under FDI attack on the communication link from $\mathcal{S}^1$ to $\mathcal{S}^2$ is studied, i.e., when $\mathcal{S}^2$ receives contaminated data $\tilde y_t^1$. During these simulations, the FDI attack signal was the chirp signal defined in the previous scenario. Fig. \ref{fig81} illustrates a comparison of the trajectory of the mobile robot formation under an FDI attack on $y_t^{[1,2]}$, both without compensation and with the application of the proposed framework. As depicted in Fig. \ref{fig81} (a), following the initiation of the FDI attack, $\mathcal{S}^2$ experiences a deviation from its intended trajectory. Consequently, subsystems $\mathcal{S}^3$ and $\mathcal{S}^4$ also deviate from their designated trajectories, as they rely on the position data received from $\mathcal{S}^2$. However, in Fig. \ref{fig81} (b), all subsystems closely adhere to their intended trajectories, highlighting the efficacy of the proposed framework. 

\subsection{Other Design Considerations}
To further investigate the effectiveness of the proposed framework, as summarized in Algorithms \ref{alg4} to \ref{alg3}, the impact of varying several design factors on the performance is examined.

\subsubsection{Effect of Overlap Between Subsequent Frames ($L$)}
Varying the overlap $L$ between subsequent segments of data influences the temporal resolution of the scalo-temproal feature extraction process. In this part, a comparative study of different values of overlap parameter $L$ is provided, where the overlap value is selected from the set $L=\{15,30,45,54,59\}$, corresponding to 25, 50, 75, 90, and 99 percent overlap between subsequent segments of data, respectively. One design aspect is that, since a prediction is needed at each time step, the base model chosen by the complexity classifier CNN is used to predict the time series data over $L$ consecutive time steps. To compare the performance of the proposed framework with different overlap values, the RMSE metric between the predicted sequence and the test data points as well as the average inference time of each case were calculated, where the average inference time of the proposed framework, denoted by $T_{infr}$, is calculated using the following equation:
\begin{equation} \label{eq4.11}
   T_{infr}=\frac{1}{N_I}\sum_{i=1}^{N_I} T_{infr}^i
\end{equation}
where $N_I$ is the number of segments of test data, and $T_{infr}^i$ is the inference time for $i^{\text{th}}$ segment of test data, which is calculated as follows:
\begin{equation} \label{eq4.12}
    T_{infr}^i =\frac{1}{M-L}\left [ T_f^i+T_c^i+(M-L)T_p^i \right ] 
\end{equation}
where $M$ is the frame length, ans $T_f^i,\ T_c^i,\ T_p^i$ are the processing times for generating multi-channel scalogram image frame from $i^{\text{th}}$ segment of test data, classifying its complexity, and performing time series forecasting using the selected model, respectively. Table \ref{tab4.8} shows the comparison results for different overlap values.
\begin{table}[t]
	\caption{Comparison of RMSE metric and inference time on test dataset using different overlap values}
	\begin{center}
	\label{tab4.8}
\begin{tabular}{|c|c|c|c|}
	\hline
	  $L$ & $N_I$ & RMSE & Average Inference Time\\
	\hline
      15 & 33 & 0.330 m & 4.57 ms\\
 
      30 & 49 &  0.352 m & 5.32 ms\\
      
      45 & 97 & 0.328 m & 7.21 ms\\   
      
      54 & 241 & 0.287 m & 12.61 ms\\  
      
      59 & 1442 & 0.279 m & 63.26 ms\\  
	\hline
\end{tabular}
\end{center}
\end{table}

As demonstrated in Table \ref{tab4.8}, increasing the overlap generally improves prediction accuracy. However, this also increases the computational load, which is undesirable for real-time data processing. Therefore, in this paper, the overlap parameter was set to $L=54$, which does not significantly degrade the performance compared to $L=59$, but leads to $80.07\%$ reduction in the average inference time.



\subsubsection{Removal of Stage I}
In this part, the impact of removing Stage I from the proposed framework on prediction performance is analyzed. Removing Stage I offers a trade-off by reducing the computational burden through the direct input of time series data into the base models, eliminating the need for dynamic model selection. However, this comes at the cost of simultaneously utilizing all base models, which ultimately increases the overall computational load of the prediction algorithm. Having Stage I removed from the proposed framework, the output of the stacked ensemble learning architecture is calculated by averaging across all base models, as expressed in the following equation, which is similar to the framework proposed in \cite{WANG2022100542}:
\begin{equation} \label{eq4.13}
    \hat a^{\text{stacked}}=\frac{1}{n}\sum_{l=1}^{n}\hat a^l
\end{equation}

To compare the performance of proposed framework with the predictions from (\ref{eq4.13}), the RMSE metric between the predicted sequence and the test data points as well as the average inference time of each framework were calculated. The comparison results are summarized in Table \ref{tab4.7}.
 \begin{table}[t]
	\caption{Comparison of RMSE metric and inference time on test dataset using average of base models against the proposed framework}
	\begin{center}
	\label{tab4.7}
\begin{tabular}{|c|c|c|}
	\hline
	 Model & RMSE & Average Inference Time \\
	\hline
      Framework \cite{WANG2022100542}, $n=3$ & 0.414 m & 13.28 ms\\
 
      Framework \cite{WANG2022100542}, $n=5$ & 0.339 m & 23.31 ms\\
      
      Proposed Framework & 0.287 m & 12.61 ms \\    
	\hline
\end{tabular}
\end{center}
\end{table}

Table \ref{tab4.7} shows that the proposed framework has $15.34\%$ less prediction error than calculating the average of all base models, while being $45.90\%$ faster. A key consideration is that the inference time of the prediction algorithm increases with the number of base models, $n$, when using the average prediction, leading to scalability challenges. The proposed framework addresses this challenge by selecting only one base model from the stack for prediction at each time step. To verify the effect of number of base models $n$ on the performance of the framework, a subset of three models from the stacked ensemble learning architecture, including GRU, CNN-LSTM, and Modified Xception-LSTM were used to perform prediction based on (\ref{eq4.13}), and the results are summarized in the first row of Table \ref{tab4.7}. These results justify the use of the proposed framework in this paper, as it not only significantly reduces prediction error by 30.68\%, thereby improving the accuracy of the prediction algorithm, but also increases computational efficiency by reducing inference time by 5.05\%, demonstrating that the framework can effectively balance the trade-off between accuracy and speed, which is crucial for real-time applications where both accurate and prompt predictions are essential.

\subsection{Comparison with Related Data-Driven Frameworks in the Literature}
As discussed in section \ref{sec1}, the proposed framework addresses the shortcomings of the existing data-driven approaches for mitigating FDI attacks. Among the frameworks outlined in Table \ref{tab3}, the frameworks in \cite{9352502} and \cite{RAGHUVAMSI2023112565} demonstrate the closest resemblance to the one presented in this paper, as they are designed to operate on a single sensor within the system, rather than approximating the dynamics of the entire system. Additionally, two widely-used ensemble learning methods, Bagging and XGBoost, were implemented to demonstrate the superiority of the proposed framework compared to other state-of-the-art methods. The frameworks were evaluated by comparing the RMSE between the actual and predicted time series data, as well as their average inference time. A summary of these comparisons is provided in Table \ref{tab4.9}.
\begin{table}[t]
	\caption{Comparison of the proposed framework against existing literature and state-of-the-art ensemble learning models}
	\begin{center}
	\label{tab4.9}
\begin{tabular}{|c|c|c|}
	\hline
	 Model & RMSE & Average Inference Time \\
	\hline
      Framework \cite{9352502} & 1.193 m & 2.56 ms\\

      Framework \cite{RAGHUVAMSI2023112565} & 1.556 m & 3.59 ms\\
 
      Bagging & 1.346 m & 58.98 ms\\
      
      XGBoost & 1.483 m & 11.65 ms\\
      
      Proposed Framework & 0.287 m & 12.61 ms \\    
	\hline
\end{tabular}
\end{center}
\end{table}

As shown in Table \ref{tab4.9}, the proposed framework achieves superior performance compared to all other frameworks in terms of RMSE, showing improvements of 75.94\%, 81.57\%, 78.68\%, and 80.65\% over Framework \cite{9352502}, Framework \cite{RAGHUVAMSI2023112565}, Bagging, and XGBoost, respectively. While the proposed framework is faster than Bagging, it is slower than Framework \cite{9352502}, Framework \cite{RAGHUVAMSI2023112565}, and XGBoost. Nevertheless, given the sampling period of the mobile robot used as the case study in this paper, $T_s=50\ ms$, the proposed framework is sufficiently fast to perform real-time FDI attack mitigation, making it a viable choice for this application.

\section{Conclusion} \label{sec5}
In this paper, a data-driven framework for online mitigation of FDI attacks on sensor measurements of networked control systems was presented. The proposed framework uses a two-stage stacked ensemble learning architecture. The first stage performs meta learning through a time-frequency domain analysis of the raw time series data which reduces the computational burden of the proposed framework. In the second stage, the selected model among a stack of deep neural networks is used for time series forecasting and providing a $q$-steps ahead prediction of time series data. The stacked architecture is comprised of five networks, namely LSTM, GRU, CNN-LSTM, ResNet-50-LSTM, and a modified version of Xception network succeeded with a LSTM network, referred to as Modified Xception-LSTM. Then, the predicted time series is used to remove malicious data from sensor measurements in an online manner, provide an uncontaminated feedback signal for controller, and avoid any undesired shutdown to system. Numeric simulations verify the effectiveness of the proposed framework.

Even though the proposed framework has demonstrated promising results in mitigating FDI attacks on sensor measurements of networked control systems, there are still some interesting topics which are worth being investigated and studied. Some of the possible future research directions are as follows:
\begin{itemize}
	\item Leveraging predictive control schemes to utilize data-driven methods' capability in forecasting future value of time series and study the closed-loop performance of the proposed framework with $q>1$
    \item Studying dynamic selection of more than one base time series forecasting model for FDI mitigation
    \item Studying the case of simultaneous FDIs on actuator commands and sensor measurements
    \item Adapting the proposed framework to mitigate more complex attack scenarios, such as hybrid cyberattacks \cite{9980533}
\end{itemize}


\bibliographystyle{ieeetr}
\bibliography{references}

\end{document}